\documentclass[preprint]{iucr}              

     \journalcode{S}              

\begin{document}                  


        \title{Analysis of long-lived effects in high-repetition-rate stroboscopic transient X-ray absorption experiments on thin films}


        \aff[a]{Faculty of Physics and Center for Nanointegration Duisburg-Essen (CENIDE), University of Duisburg-Essen, Lotharstr.~1, 47057 Duisburg, Germany}
        \aff[b]{European XFEL, Holzkoppel~4, 22869 Schenefeld, Germany}
        \aff[c]{Elettra Sincrotrone Trieste, Strada Statale~14 - km~163,5, 34149 Basovizza, Trieste, ITALY}
        \aff[d]{Universit\'{e} de Strasbourg, CNRS, Institut de Physique et Chimie des Mat\'{e}riaux de Strasbourg, UMR~7504, 67000 Strasbourg, France}
        \aff[e]{Dipartimento di Elettronica, Informazione e Bioingegneria, Politecnico di Milano, 20133 Milano, Italy}
        \aff[f]{Istituto Nazionale di Fisica Nucleare, Sez., Milano, 20133 Milano, Italy}
        \aff[g]{Deutsches Elektronen-Synchrotron DESY, Notkestr.~85, 22607 Hamburg, Germany}
        \aff[h]{EXTOLL GmbH, 68159 Mannheim, Germany}
        \aff[i]{Institute for Computer Engineering, University of Heidelberg, Im Neuenheimer Feld~368, 69120 Heidelberg, Germany}
        \aff[j]{Max-Planck Institute for the Structure and Dynamics of Matter, Luruper Chaussee~149, 22761 Hamburg, Germany}
        \aff[k]{University of Southampton, Southampton SO17 1BJ, United Kingdom}
        \aff[l]{Dipartimento di Elettronica, Informazione e Bioingegneria, Politecnico di Milano, 20133 Milano, Italy}
        \aff[m]{Istituto Nazionale di Fisica Nucleare, Sez., Milano, 20133 Milano, Italy}
        \aff[n]{Materials Chemistry and Catalysis (MCC), Debye Institute for Nanomaterials Science, Utrecht University, Universiteitslaan~99, 3584 CG, Utrecht, The Netherlands}
        \aff[o]{Department of Molecular Sciences and Nanosystems, Ca’ Foscari University of Venice, 30172 Venezia, Italy}
        \aff[p]{Laboratori Nazionali di Frascati, INFN, Via Enrico Fermi 54, 00044 Frascati (Roma), Italy}
        \aff[q]{Paul Scherrer Institut, Forschungsstr.~111, 5232 Villigen PSI, Switzerland}
        \aff[r]{Department of Materials, ETH Zurich, 8093 Zurich, Switzerland}
        \aff[s]{Institute for Electric Power Systems, University of Applied Sciences and Arts Northwestern Switzerland, 5210 Windisch, Switzerland}
        \aff[t]{MAX IV Laboratory, Lund University, Box~118, 221 00 Lund, Sweden}
        \aff[u]{Department of Physics, AlbaNova University Center, Stockholm University, SE-10691 Stockholm, Sweden}
        \aff[v]{Institute for Solid State Physics, The University of Tokyo, Kashiwa, Chiba 277-8581, Japan} 

        
        \cauthor[a]{Tobias}{Lojewski}{tobias.lojewski@uni-due.de}
        
        \author[b]{Lo\"ic}{Le Guyader}
        \author[b,c]{Naman}{Agarwal}
        \author[d]{Christine}{Boeglin}
        \author[b]{Robert}{Carley}
        \author[e,f]{Andrea}{Castoldi}
        \author[b]{Carsten}{Deiter}
        \author[g]{Robin}{Y. Engel}
        \author[h,i]{Florian}{Erdinger}
        \author[b,j,k]{Hans}{Fangohr}
        \author[l,m]{Carlo}{Fiorini}
        \author[b]{Natalia}{Gerasimova}
        \author[b]{Rafael}{Gort}
        \author[n]{Frank}{de Groot}
        \author[b]{Karsten}{Hansen}
        \author[b]{Steffen}{Hauf}
        \author[b]{David}{Hickin}
        \author[b]{Manuel}{Izquierdo}
        \author[a]{Lea}{K\"ammerer}
        \author[b]{Benjamin}{E. Van Kuiken}
        \author[b]{David}{Lomidze}
        \author[g]{Stefano}{Maffessanti}
        \author[b]{Laurent}{Mercadier}
        \author[b]{Giuseppe}{Mercurio}
        \author[g]{Piter}{S. Miedema}
        \author[d]{Matthias}{Pace}
        \author[b,o]{Matteo}{Porro}
        \author[p]{Javad}{Rezvani}
        \author[a]{Nico}{Rothenbach}
        \author[q]{Benedikt}{R\"osner}
        \author[b,g]{Andrey}{Samartsev}
        \author[b]{Justina}{Schlappa}
        \author[r,s]{Christian}{Stamm}
        \author[b]{Martin}{Teichmann}
        \author[b]{Monica}{Turcato}
        \author[b,t]{Alexander}{Yaroslavtsev}
        \author[q]{Florian}{D\"{o}ring}
        \author[b]{Andreas}{Scherz}
        \author[q]{Christian}{David}
        \author[g,u]{Martin}{Beye}
        \author[a,v]{Uwe}{Bovensiepen}
        \author[a]{Heiko}{Wende}
        \author[a]{Andrea}{Eschenlohr}
        \author[a]{Katharina}{Ollefs}

     \maketitle                        
     

        \begin{synopsis}
        We disentangle and analyze the secondary contribution of long-lived excitations that occur in high-repetition-rate stroboscopic X-ray absorption spectroscopy experiments of Ni and NiO thin film samples. 
        \end{synopsis}


        \begin{abstract}
        Time-resolved X-ray absorption spectroscopy (tr-XAS) has been shown to be a versatile measurement technique for investigating non-equilibrium dynamics. Novel X-ray free electron laser (XFEL) facilities like the European XFEL offer increased repetition rates for stroboscopic XAS experiments through a burst operation mode, which enables measurements with up to 4.5~MHz. These higher repetition rates lead to higher data acquisition rates but can also introduce long-lived excitations that persist and thus build up during each burst. Here, we report on such long-lived effects in Ni and NiO thin film samples that were measured at the European XFEL. We disentangle the long-lived excitations from the initial pump-induced change and perform a detailed modelling-based analysis of how they modify transient X-ray spectra. As a result, we link the long-lived effects in Ni to a local temperature increase, as well as the effects in NiO to excited charge carrier trapping through polaron formation. In addition, we present possible correction methods, as well as discuss ways in which the effects of these long-lived excitations could be minimized for future time-resolved X-ray absorption spectroscopy measurements. 
        \end{abstract}


        \section{Introduction}
        
            X-ray absorption spectroscopy (XAS) is a powerful measurement technique to examine the electronic and magnetic properties of condensed matter, providing element-specific information about the electronic structure, magnetic properties, and the chemical environment \cite{stohr_magnetism_2006,philip_willmott_introduction_2019}. Due to the sensitivity to changes in the electronic structure and its element-specificity, XAS has been proven in recent years to be a great tool for investigating non-equilibrium states in condensed matter. Following an optical laser excitation of the sample, transient XAS can trace modifications of the electronic structure in time, as initially shown by Stamm et al. \cite{stamm_femtosecond_2007}. Since then, femtosecond time-resolved soft X-ray absorption spectroscopy on thin-film samples has been used to investigate changes in the magnetization \cite{stamm_femtosecond_2007,stamm_femtosecond_2010,lopez-flores_time-resolved_2012,lopez-flores_role_2013,hennecke_angular_2019}, the role of the lattice in the non-equilibrium energy transfer \cite{rothenbach_microscopic_2019,rothenbach_effect_2021}, photoexcitations in functional photocatalytic materials \cite{uemura_femtosecond_2021,uemura_hole_2022}, and optically induced spin currents \cite{eschenlohr_ultrafast_2013,eschenlohr_spin_2017,stamm_x-ray_2020,eschenlohr_spin_2021}.
            
\begin{figure}
                    \caption{(a) Exemplary X-ray absorption spectra for the NiO O K-edge (left), NiO  L$_{2,3}$-edges (middle) and Ni-metal L$_{2,3}$-edges (right). The grey line shows the ground state absorption spectrum while the coloured line shows the excited state spectrum after 0.5 ps for NiO and 0.4 ps for Ni, after 4 accumulated laser pulses per train. The coloured dots show the excited state spectrum for the same delay using 16 accumulated laser pulses. (b) Resulting pump-induced change going from 4 to 16 accumulated laser pulses per train for the three absorption edges. (c) Absolute area encompassed by the pump-induced change as a function of an increasing number of accumulated laser pulses. The lines are a guide to the eye.}
                    \includegraphics[width=1\textwidth]{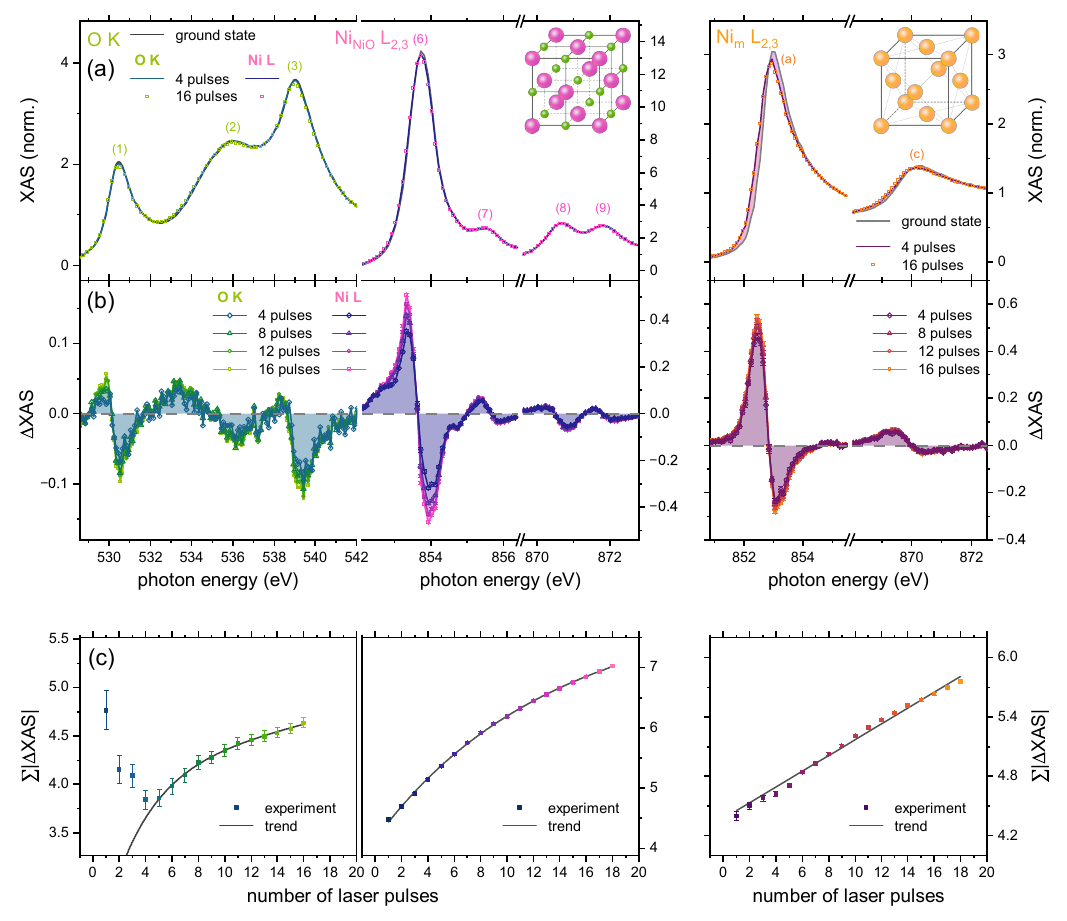}
                    \label{fig_1}
                \end{figure}

            With transient X-ray absorption being a well-established and versatile measurement technique, the focus was directed towards improving the experimental setups in regard to energy-, time resolution, and measurement time. For stroboscopic-type experiments such as pump-probe XAS, the measurement time can be reduced by increasing the brilliance of the X-ray sources as well as the repetition rate, while the time resolution can be improved up to a certain point with short X-ray and laser pulses and a more stable temporal overlap. At third-generation synchrotron radiation facilities, pulse lengths are typically between 50 and 100~ps \cite{stepanov_short_2016}. Using a laser-slicing-based technique, where only a small slice of electrons from the main electron bunch are modulated in energy, it is possible to reduce the pulse length to around 100~fs, but at the cost of the number of X-ray photons \cite{stepanov_short_2016,holldack_femtospex_2014}. X-ray free electron lasers, on the other hand, offer inherently short X-ray flashes at around 100~fs with extremely high peak brilliances. However, facilities with a normal-conducting accelerator design have historically had reduced repetition rates of typically between 60 and 120~Hz \cite{stepanov_short_2016,tschentscher_photon_2017,emma_first_2010,ishikawa_compact_2012,ko_construction_2017,patterson_coherent_2010} compared to third-generation synchrotron sources. Facilities that employ a superconducting accelerator design \cite{decking_mhz-repetition-rate_2020} are able to achieve comparatively higher repetition rates, such as FLASH, which can operate between 10~Hz and 5~kHz \cite{ackermann_operation_2007,allaria_flash2020_2021}, and the European XFEL which can reach up to 4.5~MHz  \cite{stepanov_short_2016,tschentscher_photon_2017,le_guyader_photon-shot-noise-limited_2023}. Typically, these repetition rates are achieved through a burst operation mode, with the exception of the currently ongoing LCLS-II upgrade, which is planning to achieve a 1~MHz repetition rate in continuous operation \cite{raubenheimer_lcls-ii-he_2018,brachmann_lcls-ii_2019}. Specifically at the European XFEL, the repetition rate of 4.5~MHz is achieved in a burst operation mode where up to 2700 electron bunches are generated and accelerated in short bursts (intra-burst temporal spacing of 222~ns), with a burst repetition rate of 10~Hz. With these high repetition rates and the high peak brilliance, this setup enables extremely efficient data acquisition, which allows the focus to shift towards performing a time-resolved spectroscopic analysis of the pump-induced changes to disentangle the initial excitation and subsequent relaxation dynamics \cite{lojewski_interplay_2023,lojewski_photo-induced_2024}. However, high-repetition-rate stroboscopic experiments also bring additional challenges. Long-lived excitations, in this context defined as excitations that do not fully relax back to the initial ground state before the next optical excitation, can lead to a gradual modification of the pump-induced change over the course of the burst through the additional energy that is deposited with each subsequent excitation. One such effect is laser-induced heating, which was already observed in transient XAS experiments of Ni-metal on timescales of 100 ps \cite{kachel_transient_2009}, and which we again observed in our experiments on capped thin Ni layers \cite{le_guyader_photon-shot-noise-limited_2023,lojewski_interplay_2023}. These laser-induced heating effects are a result of sample design, where the measurement geometry requires thin films on freestanding X-ray transparent substrates. Thus, the available volume allowing for heat dissipation is too small, so the heat load introduced by the laser pulse can not dissipate quickly enough. With the bunch-train X-ray delivery scheme of the European XFEL, we could analyze the pump-induced changes as a function of the accumulated number of laser pulses per train \cite{le_guyader_photon-shot-noise-limited_2023}, which is shown on the right of figure~\ref{fig_1} panel (a-c). We found that the induced changes increase linearly with the number of laser pulses, shown in panel (c) of figure~\ref{fig_1}, as each laser pulse leads to a small local increase in the sample temperature. 
            
            However, laser-induced heating is not the only long-lived effect that can occur. It has been observed that the electron trapping mechanism can influence the dynamics on long-time scales as well. For example, Foglia et al. showed that defect-mediated exciton trapping can lead to excitations which persist on a ns time scale in oxides \cite{foglia_revealing_2019}. Furthermore, Santomauro et al. showed with transient X-ray absorption spectroscopy long-lived excitations in inorganic perovskites following an optical excitation that persists up to hundreds of ns and which they link to the formation of polarons \cite{santomauro_localized_2017}. In our NiO measurements, we also find a contribution of some long-lived excitation, which appears distinctly different from the one in Ni. Instead of a linear dependence of the pump-induced change on the accumulated number of laser pulses we find in some part a saturation-like behaviour. Charge carrier trapping through polaron formation might be a possible origin of these distinctly different long-lived effects that can be observed in NiO, especially as polaron trapping has been previously observed in NiO \cite{biswas_ultrafast_2018}, and it was shown that polaron trapping can lead to extremely long-lived excited states \cite{santomauro_localized_2017}. In this work, we disentangle the contributions of the long-lived excitations from the initial pump-induced change and analyze them in regard to time delay-, fluence dependence and their modification of the spectral shape.    
            
        \section{Experimental detail}
        
            \subsection{Sample preparation}
            The Ni-metal and NiO samples were grown using molecular beam epitaxy (MBE) on 200~nm thick Si$_3$N$_4$ membranes in a 5~x~5 and 12~x~12 grid formation for Ni and NiO, respectively. For both samples, a 100~nm thick Copper layer was deposited on the back of the substrates to be used as a heatsink and to increase stability. During the growth, some rows were masked so that the resulting bare Si$_3$N$_4$ membranes can be used in combination with the beam-splitting off-axis zone plate setup \cite{le_guyader_photon-shot-noise-limited_2023} to obtain a reference signal. In the case of the Ni samples, a 20$\pm$0.7~nm thick fcc Ni layer was deposited on the uncovered membranes, which was later capped using a 2~nm MgO layer to avoid oxidation. For the NiO samples, a 37.4$\pm$0.1~nm thick NiO layer was grown by Ni evaporation under an oxygen atmosphere of 1$\cdot$10$^{-6}$~mbar. The samples were characterized using XRD and vibrating sample magnetometry to guarantee the structural and magnetic properties of the Ni and NiO samples \cite{lojewski_interplay_2023,lojewski_photo-induced_2024}. For additional details on the sample preparation, see the supplementary material of \cite{lojewski_interplay_2023,lojewski_photo-induced_2024}.
        
            \subsection{Experimental setup}
            All experimental XAS data was measured at room temperature in a stroboscopic pump-probe experiment at the Spectroscopy and Coherent Scattering Instrument (SCS) of the European XFEL \cite{tschentscher_photon_2017,le_guyader_photon-shot-noise-limited_2023,lojewski_interplay_2023}. 
            
            The Ni-metal samples are excited using an ultrashort (35~fs) optical laser pulse \cite{pergament_versatile_2016} with a photon energy of $h\nu$~=~1.5~eV (800~nm) and an incidence fluence of 12~mJ/cm$^2$. The laser pulse causes excitations of the 3$d$ electrons into unoccupied 4$sp$ states that subsequently scatter and relax to an elevated electronic temperature. The NiO samples are resonantly excited above the optical bandgap using a photon energy of $h\nu$~=~4.7~eV (266~nm) with an incidence fluence of 0.8 or 4.0~mJ/cm$^2$ to excite electrons from the O 2$p$ states into unoccupied states of the upper Hubbard band \cite{gillmeister_ultrafast_2020}. The pump pulses are synchronized with X-ray pulses to probe the modification following the laser excitation after a time delay of $\Delta t$. We measure the Ni L$_{2,3}$-edges for both Ni-metal and NiO using monochromatic \cite{gerasimova_soft_2022} linearly polarised X-ray with energies tuned between 840 and 880~eV to excite 2$p_{3/2}$ or 2$p_{1/2}$ core electrons into unoccupied 3$d$ final states. In addition, for NiO, the O K-edge is also measured using X-ray energies between 525 and 565~eV to excite 1$s$ core electrons into unoccupied 2$p$ states.
            
            All measurements were performed with the beam-splitting off-axis zone plate (BOZ) setup \cite{doring_multifocus_2020,le_guyader_photon-shot-noise-limited_2023} of the SCS instrument. The zone plate focuses and splits the incoming X-rays into three distinct beams according to the 0$^{th}$, -1$^{st}$ and 1$^{st}$ diffraction order, which allows the measurement of three spatially separated regions simultaneously. For transient pump-probe XAS measurements, this setup is used to measure the ground-, excited- state and reference signal concurrently. All measurements were performed in transmission geometry by detecting the X-ray intensity following transmission through the sample with a fast 1~Mpixel X-ray imager (DSSC camera) \cite{porro_minisdd-based_2021}. The X-ray delivery for the experiments followed the bunch-train scheme \cite{tschentscher_photon_2017,le_guyader_photon-shot-noise-limited_2023}. So, the X-ray absorption spectra were measured using bursts (trains) of high repetition rate X-ray pulses with an intra-train repetition rate of 70~kHz (14.3~$\mu$s) or 57.5~kHz (17.4~$\mu$s) for Ni or NiO, respectively. The burst (trains) are delivered with a train repetition rate of 10~Hz. The bunch-train delivery pattern was adjusted between experiments so that for the Ni metal measurements, a total of 25 pump-probe synchronized pulses were used, which was reduced to 18 synchronized pulses when measuring the NiO Ni L-edges with a pump fluence of 4~mJ/cm$^2$. The pattern was further changed for the measurements of the oxygen K-edge and the NiO Ni L-edges with a pump fluence of 0.8~mJ/cm$^2$ by dropping the first and last X-ray pulse, leaving only 16 pump-probe synchronized pulses. In this analysis, we only use the first 18 pulses from the Ni-metal measurements, but we have previously shown with another dataset that the linearly increasing behaviour with the number of laser pulses, as highlighted by figure \ref{fig_1} c), remains the same even for more accumulated laser pulses \cite{le_guyader_photon-shot-noise-limited_2023}.
            
            The XAS data is initially processed and evaluated with the dedicated SCS toolbox \cite{le_guyader_photon-shot-noise-limited_2023,fangohr_data_2020,noauthor_scs_2024} by calculating the negative logarithm of the transmitted ground- or excited-state signal, divided by the reference signal. Additionally, the flat field and non-linearity correction detailed in \cite{le_guyader_photon-shot-noise-limited_2023} are applied. The toolbox also allows for a separation of the data according to the accumulated number of laser pulses within each train, which we use to characterize the secondary long-lived effects. Following this initial processing, the XAS data is exported for each spectrum and optical laser pulse for further evaluation, which is described in the following section. The raw data of all experiments shown in this publication is publicly available. For details, see section \ref{Data_availability} or \cite{noauthor_Ni_Data_2020,noauthor_NiO_Data_2021}.

        \section{Data processing}
        
            As we aim to disentangle and discuss the primary contribution from the pump-induced changes and the secondary contribution building up through long-lived excited states, we have to ensure comparability between the XAS spectra, especially with changes in the experimental setup between experiments. To that end, we perform a background correction and normalize the spectra to the invariant edge jump \cite{wang_l-edge_2018}, as we will detail in the following section. Additionally, we can apply the modelling approach that we previously used to great success in modelling the induced change itself \cite{lojewski_interplay_2023,lojewski_photo-induced_2024} to also model the secondary contribution of long-lived effects.

            \subsection{Background correction and normalization}
\begin{figure}
                    \caption{(a) Exemplary background correction of a NiO L$_{2,3}$ ground and excited state spectrum [$\Delta$t = 0.5~ps] (left), as well as the pump-induced change (right). The linear background (green solid line) is constructed from four points (orange squared) describing the average value at the beginning and end of the pre- and post-edge region (right). (b) The absorption spectra are normalized so that the pre-edge (blue area) corresponds to zero and the post-edge (red area) corresponds to one. (c) Final normalized spectra (left) and corresponding pump-induced change (right).}
                    \includegraphics[width=8.85cm]{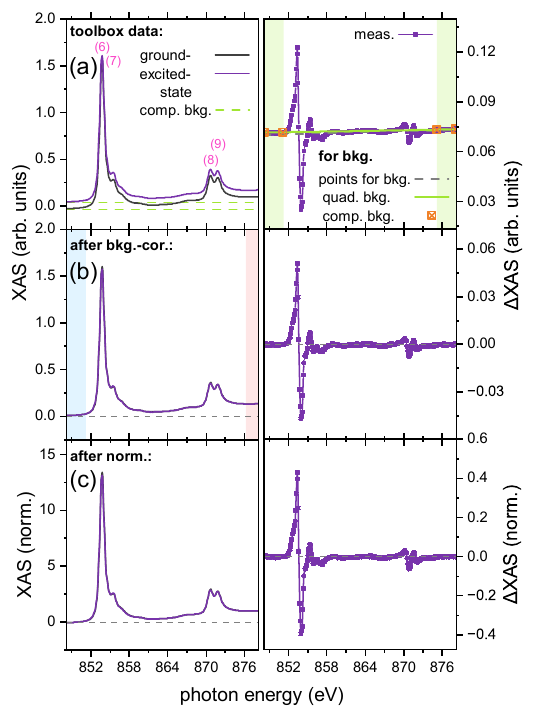}
                    \label{fig_2}
                \end{figure}

            One exemplary NiO Ni L$_{2,3}$ - absorption spectrum ($\Delta$t~=~0.5~ps), as obtained from the toolbox evaluation, is shown in panel (a) of figure~\ref{fig_2} on the left with the corresponding pump-induced change ($\Delta$XAS) on the right. Between the ground and excited state spectra, a clear offset is visible, which we attribute to some minor differences in the -1$^{st}$ and 1$^{st}$ order beams split by the zone plate. This difference is also visible in the $\Delta $XAS as an offset. In addition, the spectra diverge slightly with increasing X-ray energy, which could also be linked to the zone plate setup. Both of these effects are corrected using a background correction. For the background correction, we take two regions where no pump-induced changes are visible, which are the L$_{3}$ pre-edge (848.2 - 851.2~eV) and L$_{2}$ post-edge region (875.2 - 878.2~eV), highlighted by the green shading in figure~\ref{fig_2}. As there is a slightly increasing slope in these regions, we fit two linear functions, shown as dashed grey lines in figure~\ref{fig_2}, to obtain a mean value for the beginning and end of each region, respectively, which is displayed as orange data points. Using these four mean values, we can construct the linear background shown in figure~\ref{fig_2} as the green line. The $\Delta $XAS is then corrected by subtracting the background, while the spectra are corrected by subtracting two linear functions with half the slope and opposite signs pulling them on top of one another. The results of this background correction are shown in panel (b) of figure~\ref{fig_2}. Following the correction, the ground and excited state spectra coincide perfectly in the pre and post-edge region, and the $\Delta $XAS also shows zero induced changes in these regions. The spectra are then normalized so that the mean L$_{3}$ pre-edge region, highlighted by the blue shading in panel (b) of figure~\ref{fig_2}, corresponds to zero, while the L$_{2}$ post-edge region, which corresponds to the invariant edge jump and is marked by the red shading, corresponds to one. The $\Delta $XAS is normalized accordingly, and the final results are shown in panel (c) of figure~\ref{fig_2}.
            
            The background correction and normalization of the Ni metal Ni L$_{2,3}$ - absorption edge is done in exactly the same way and is, as such, not further shown.
            
\begin{figure}
                    \caption{(a) Exemplary background correction of a NiO oxygen K-edge ground and excited state spectra [$\Delta$T = 0.5~ps] (left), as well as the pump-induced change (right). A quadratic background (dashed grey line) is constructed by fitting the $\Delta$XAS in the green highlighted regions (orange squared). (b) Additional background contribution that results from a static difference between the two membranes, which is corrected using the total background (pink solid line). (c) The absorption spectra are normalized so that the pre-edge (blue area) corresponds to zero and the post-edge (red area) corresponds to one. (d) Final normalized spectra (left) and corresponding pump-induced change (right)}
                    \includegraphics[width=8.85cm]{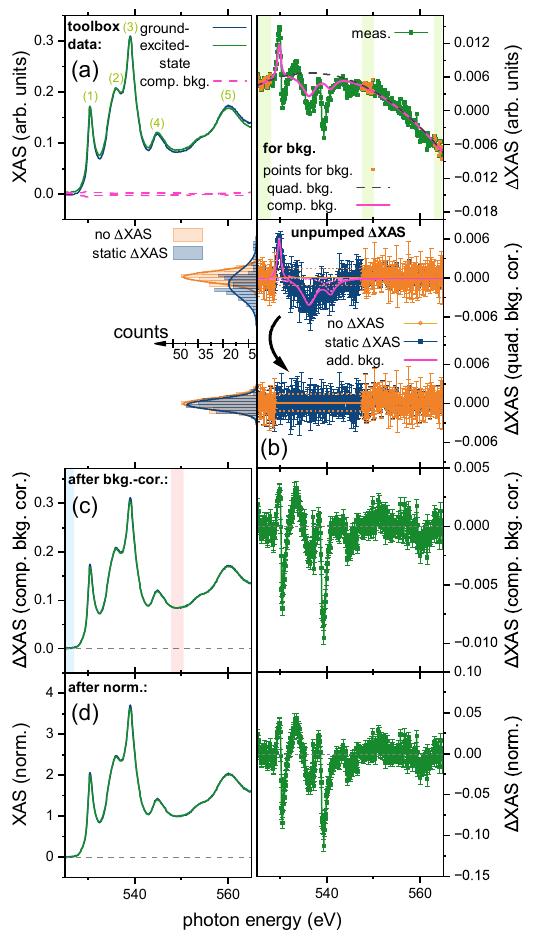}
                    \label{fig_3}
                \end{figure}

            For the oxygen K-edge, the ground and excited state spectra also show a slight offset like the Ni L$_{2,3}$ - edge, but in addition, they no longer diverge linearly with the X-ray energy but rather show a quadratic dependence, as highlighted in figure \ref{fig_3} a) with the grey dashed line. As such, the linear approximation of the background is no longer possible, and a quadratic function has to be used instead. In order to obtain a good approximation of the background with the quadratic function, we choose, in addition to the pre- (525-528~eV) and post-edge regions (563 - 564.8~eV), an additional region between the fourth and fifth fine-structure peak (547.5 - 550~eV). From our modelling, as described later in this section, we expect this region to show no induced changes that would be the result of an energy shift or broadening. The three regions used as reference for the background correction are marked with green shading in panel (a) of Figure~\ref{fig_3}. Instead of determining two mean values per region as done before with the Ni L - edges, we find better results by using all data points in this region directly to fit the quadratic function. The data points used for the background correction are also highlighted in orange in panel (a) of Figure~\ref{fig_3}. The resulting quadratic background is shown in panel (a) of Figure~\ref{fig_3} as a grey dashed line. In addition we find another modification that occurs even for a static measurement. This modification is visible in the first X-ray pulse of the train even before the first laser excitation and appears between the first and fourth fine-structure peaks. This contribution is apparent immediately in the first measurement at that sample position and remains unchanged for all measurements. So, after subtracting the previously determined quadratic background for each scan, we average the $\Delta$XAS to obtain a better description of the remaining modifications. The average is shown in the top part of panel (b) in figure~\ref{fig_3}. To further differentiate the contributions from the noise, we investigate the distribution of data points in the pre- and post-edge regions (orange data points) and the region of interest between the first and fourth fine structure peaks (blue data points), which is shown to the left of panel (b). The distribution of data points in the pre- and post-edge regions shows a normal distribution, indicating that these regions include only noise and no distinguishable features. The distribution in the region of interest, however, deviates strongly from a normal distribution and appears much broader compared to the pre- and post-edge regions. To clearly separate the contributions from random noise spikes, we use the 95-5 percentile of the pre- and post-edge distribution, with any modification where three or more connected data points that lie outside of this region should be considered to be real contributions. This region is shown in panel (b) of figure~\ref{fig_3} as the dotted orange lines. Using this selection criterium, we can identify one narrower positive modification around 529.7 eV, followed by a broader negative modification between 534.5 and 543~eV. To describe these modifications, we use a multiple-curve fit with up to four Lorenz curves, which are fitted simultaneously. From the fit, we obtain a description of the modifications with three Lorentz curves, which are shown in figure~\ref{fig_3} as the dashed, dotted and dashed-dotted lines. The combination of these three contributions makes up the additional background correction, which is shown in the figure as a straight line. Using this correction, we can fully correct the modifications as shown at the bottom in panel (b) of figure~\ref{fig_3}. Following the correction, the distribution of data points exactly matches the distribution that results from the noise.
            
            At this point, we want to stress that these modifications are distinctly different from the other modifications discussed in this work and also from laser-induced sample damage, which we could observe at a different window in one scan (see appendix of \cite{lojewski_photo-induced_2024} for more details). We attribute this difference to non-completely homogeneous sample properties across the three membranes used for the measurement, which could be the result of technical limitations with our MBE chamber. For one, the evaporation cells are slightly offset in regards to the sample normal and, as such, evaporate at an angle. This can lead to thickness variations of up to 2~\% over the area of the three membranes. Additionally, in our chamber, it is not possible to grow samples at high temperatures and in an oxygen atmosphere. As a result, the NiO samples were grown at room temperature, which could cause low vacancy mobility and could result in a nonuniform distribution of oxygen vacancies between the membranes.
            
            Nevertheless, this effect can be corrected using this additional background correction, which is also implemented for the rest of the oxygen K-edge spectra, resulting in the pink solid line in panel (a) of figure~\ref{fig_3}. Similar to the Ni L-edges, we correct both the ground and excited state spectra by pulling them on top of one another and subtracting the complete background from the pump-induced change. The results of this complete background correction are shown in panel (c) of figure~\ref{fig_3}. Afterwards, the spectra are normalized as was done before with the Ni L-edge spectra by setting the mean value in the pre-edge region (highlighted by blue shading) to zero and the mean value in the post-edge region (highlighted by red shading) to one. For the oxygen K-edge, we choose the post-edge region just behind the fourth fine-structure peak, as the region behind the sixth fine-structure peak is not sufficient for comparable normalization. The results of the normalization are shown in panel (d) of figure~\ref{fig_3}.
            
            \subsection{Modelling the Pump-induced Change} 
            An important part of analyzing the observed pump-induced changes is disentangling the underlying modifications of the spectrum that cause them. To accomplish this, we model two of the most common modifications observed in transient X-ray absorption spectroscopy, namely a rigid energy shift \cite{stamm_femtosecond_2007} and a spectral broadening \cite{carva_influence_2009}, and fit the results to the experimental data. This not only helps us identify the individual contributions of this energy shift and broadening but also additional modifications that cannot be described by the modelling.
            
                \begin{figure}
                    \caption{(a) Ground state spectra of the Ni-metal Ni L$_{2,3}$-edge (top), NiO Ni L$_{2,3}$-edge and oxygen K-edge (bottom) with modelled curves for no modification (solid), energy shift only (dashed line), broadening only (dotted line) and a combination of the two (dashed-dotted line) with resulting $\Delta$XAS in (b).}
                    \includegraphics[width=8.85cm]{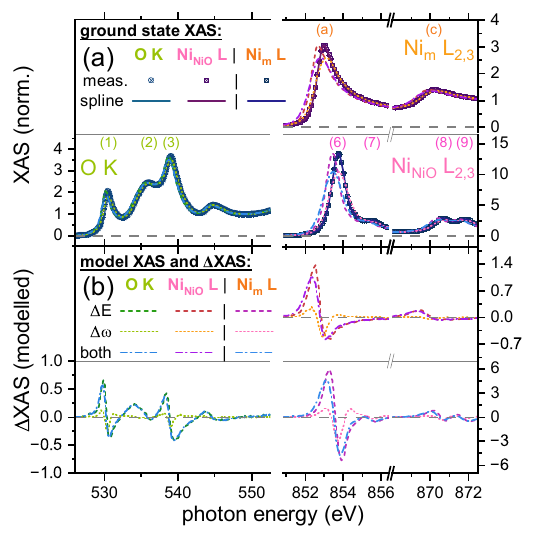}
                    \label{fig_5}
                \end{figure}
            
            To model the $\Delta$XAS, we start by calculating an Akima spline of the ground state spectrum, which allows us to freely shift the energy during the modelling by shifting the axis of the spline by the amount $\Delta$E relative to the unpumped spectrum. The broadening is modelled by convoluting the spline with a Gaussian with the full-width half maximum (FWHM) $\omega$. Exemplary modelled curves for each absorption edge are shown in figure~\ref{fig_5} (a) for an energy shift only, a broadening only and a combination of the two. 
            
            The resulting pump-induced change of these modifications is then calculated as the difference between the original ground state spectrum and the modified Akima spline. The $\Delta$XAS for each modelled curve is shown in panel (b) of figure~\ref{fig_5}. In order to describe the experiments, the modelled changes are then compared to the pump-induced changes, and the two parameters $\Delta$E and $\omega$ are adjusted to minimize the difference between the two. For the result presented here, we use the ground state spectrum with the maximum number of X-ray pulses as it offers the best data quality and does not change with the number of pulses. Furthermore, to characterize the influence of the long-lived effects, we fit a selection of curves simultaneously, which allows us to keep one parameter fixed for curves with a different number of laser pulses and optimize it globally instead. This approach offers the advantage that it is possible to try and model the effect of the long-lived states by variation of only one of the parameters, which is of great interest in disentangling the origin of these effects.
            
        \section{Results}
        
                \begin{figure}
                    \caption{(a) $\Delta$XAS for each absorption edge after 4 and 16 accumulated laser-pulses compared to the individual contribution of long-lived effects with the average contribution overlayed. (b) Average contribution of long-lived effects after 4 and 16 accumulated laser-pulses for 0.8 and 4.0~mJ/cm$^2$. (c) Average change in the absolute area encompassed by the pump-induced change as a function of an increasing number of accumulated laser pulses.}
                    \includegraphics[width=1\textwidth]{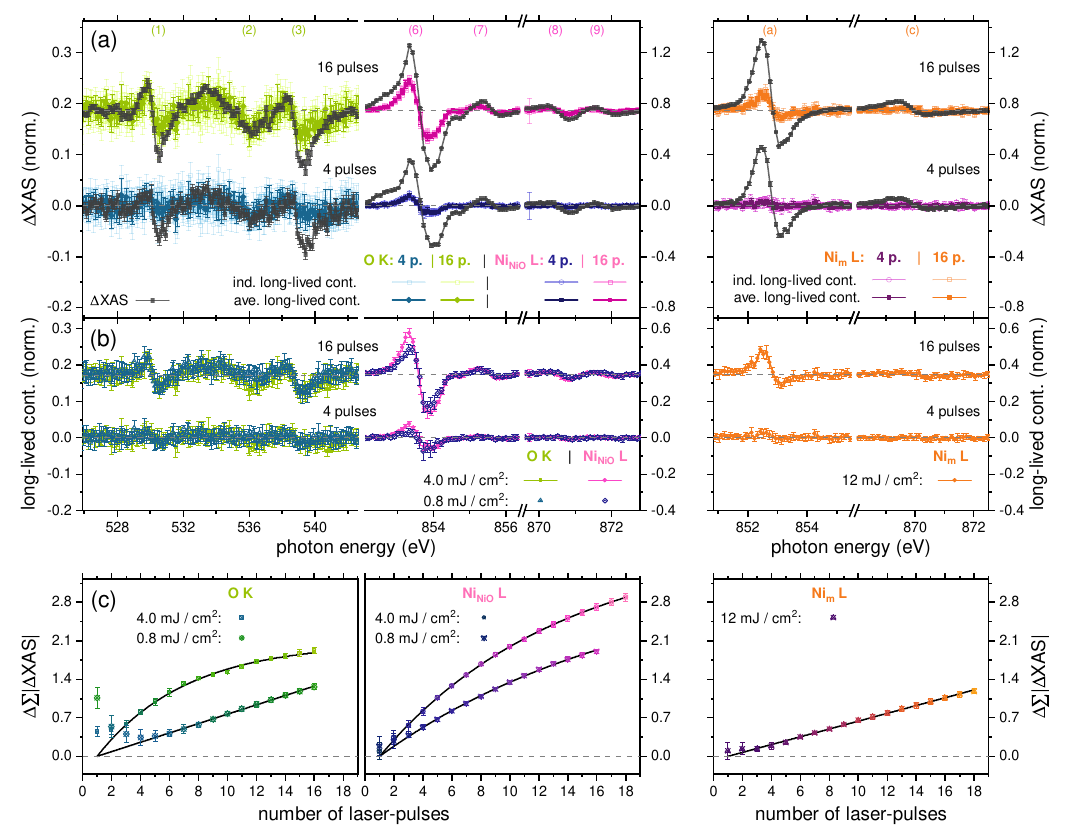}
                    \label{fig_6}
                \end{figure}
            
            In this work, we aim to separate the contribution of secondary long-lived excitations from the initial pump-induced excitations. In order to do that, we divide each measurement into sub-scans according to the number of accumulated laser pulses per train. As a result, we obtain either sixteen or eighteen ground-, excited state spectra with the corresponding $\Delta$XAS. To disentangle the individual contributions in the $\Delta$XAS, we subtract the signal measured after just the first laser pulse in each train, representing purely the initial pump-induced change, from subsequent scans, which include multiple laser excitations. Doing this allows us to investigate how the shape of the secondary contribution changes after a certain number of laser excitations per train (figure~\ref{fig_6} (a-b)) and how the area under the $\Delta$XAS evolves with the number of laser pulses (figure~\ref{fig_6} (c)) for different time delays $\Delta$t and incidence fluence. 
            
            Figure~\ref{fig_6} panel (a) shows the individual contributions of the long-lived excitations measured at different time delays $\Delta$t as semi-transparent lines in the background in comparison with an exemplary pump-induced change measured after a time delay $\Delta$t of 0.5~ps for NiO and 0.4~ps for Ni-metal. All the individual contributions show the same shape and scale of modification for all absorption edges after both four and sixteen laser excitations. This indicates that the rise time of these secondary longer-lived excitations is much longer than the initial time delay $\Delta$t of the experiment, so their contribution during the first initial laser excitations is comparably small for all measurements. To allow for a more rigorous discussion of changes in shape and scale of these long-lived contributions for different numbers of accumulated laser pulses and the two incidence fluence, the individual contributions are averaged, which is shown as the solid overlayed curve in Figure~\ref{fig_6} panel (a). 
            
            Figure~\ref{fig_6} panel (b) shows the average contribution of the long-lived excitations after four and sixteen accumulated laser excitations per train and for a pump incidence fluence of 0.8 and 4.0~mJ/cm$^2$ for NiO and 12~mJ/cm$^2$ for Ni-metal. In general, the shape of these contributions appears to follow the shape of the $\Delta$XAS itself, with only some minor differences. At the NiO Ni L-edge, the pre-edge feature around 852.6~eV, which we discuss more extensively in \cite{lojewski_photo-induced_2024}, is entirely unaffected by long-lived effects. Additionally, the NiO L$_3$-satellite (7) shows some differences compared to $\Delta$XAS. At the NiO O K-edge, the long-lived effects show a more symmetrical modification, with the differences being most apparent around the 1st and 3rd fine structure peaks (1,3). The Ni-metal L-edges show no major differences between the long-lived contribution and the $\Delta$XAS other than the intensity. Figure~\ref{fig_6} (b) also shows how the shape of the long-lived excitations evolves with the accumulated number of laser excitations per train. For all the absorption edges, the contribution of the long-lived excitations increased with an increasing number of pulses, resulting in the increase of the absolute area previously shown in figure~\ref{fig_1} panel (c). The number of pulses per train generally does not appear to influence the shape of the induced change but only increases the intensity. The incident fluence of the pump pulse does, however, affect both the shape and intensity of the induced changes, which can be seen directly at the NiO L$_3$-edge. At the NiO L$_3$-edge, the excitation with 4.0~mJ/cm$^2$ results in a stronger positive change around 853.3~eV, while the negative change at 853.9~eV is fairly similar for both 0.8 or 4.0~mJ/cm$^2$. This difference becomes more accentuated with an increasing number of laser pulses per train, as seen in figure~\ref{fig_6} (b). In contrast, the O K-edge does not display as strong a dependence on the pump fluence as the NiO L-edges, with the contribution of the long-lived excitations showing mostly the same shape for the two pump fluences. The only difference appears after sixteen laser pulses around the 2nd fine structure peak (2). 
            
            Figure~\ref{fig_6} panel (c) shows the average increase in the absolute area of the long-lived excitations. The increase of the absolute area also appears to be independent of the initial time delay $\Delta$t, showing the same trend for each absorption edge and fluence, which, as discussed before, further indicates that the contribution of these long-lived excitations is rather small on the initial timescales. However, if the noise level is high with respect to the long-lived contribution, the level of noise itself contributes significantly to the sum and causes a difference between measurements. This results in a larger error and elevated values for the first laser pulses as random noise fluctuations are the majority contributor to the sum, which can be seen at the O K-edge for 0.8~mJ/cm$^2$.

            As before, an increase in the absolute area due to the contribution of the long-lived excitations appears independent of the time delay of individual measurements so that the average of multiple measurements can be used. The only difference between measurements can be observed for the first few laser pulses where the noise level is large. Here, the contribution of summing over the noise can outweigh the contribution of the secondary effects, leading to an initially larger absolute area, which first decreases before increasing with the number of laser pulses, as seen for the O K-edge. Comparing the increasing trend for the two laser fluences further accentuates the fast increase of the long-lived contributions at the larger pump fluence. Additionally, comparing the behaviour of the contributions in the Ni-metal and NiO samples further highlights the different trends for the two samples, with a linear increase in Ni-metal and a more saturating type of behaviour in NiO. We emphasize this difference by describing the trend with the following function:
            
                \begin{math}
                    \Delta\sum\left|\Delta XAS\right|\left(n\right) = A_1 \cdot \left( 1 - exp\left( \frac{-(n-n_0)}{\theta} \right) \right) + A_2 \cdot \left( n - n_0 \right),
                \end{math}
            
            Here, the first term describes a saturating behaviour while the second term describes the linear behaviour, with $n$ as the number of accumulated laser pulses per train, $n_0$ as the first pulse and $A_{1,2}$ corresponding to the amplitude of the individual terms. The description using this function aims to separate the linear contribution, which is most likely related to a local increase in temperature from the saturating behaviour, which could be linked to other effects like polaron formation, as a saturation effect for polaron formation as a function of laser fluence has been previously observed \cite{zhang_stable_2023}. For the Ni-metal L-edges, the fit clearly shows a linear dependence with a slope of 0.071$\pm$0.001. For the NiO L$_3$ edge, the fit shows predominantly a saturating behaviour with a similar amplitude A$_1$ of 3.4$\pm$0.1 and 3.55$\pm$0.68 for 0.8 and 4.0~mJ/cm$^2$, respectively, with the number of pulses necessary to reach saturation $\theta$ decreasing from 17.83$\pm$0.65 to 11.71$\pm$1.53. A small linear contribution might be present for 4~mJ/cm$^2$, which shows a linear slope A$_2$ of 0.01$\pm$0.02. However, due to the error, a clear separation between linear and saturating contributions can not be made. For 0.8~mJ/cm$^2$ the slope of the linear contribution becomes deminishingly small. The O K-edge excited with 4.0~mJ/cm$^2$ shows a saturating behaviour with an amplitude A$_1$ of 2.07$\pm$0.04 and the number of pulses necessary to reach saturation $\theta$ being 6.16$\pm$0.27, while the slope of the linear contribution goes towards zero. The excitation with 0.8~mJ/cm$^2$ shows a mostly linear response with a slope of 0.085$\pm$0.001. However, because the noise level is so high at the oxygen K-edge compared to the relatively small contribution of both the pump-induced change and the contribution of long-lived excitations, it becomes difficult to make a conclusive statement regarding the evolution of the long-lived excitations.
            
            Overall this analysis highlights the challenge with these accumulative effects. As the contribution of these long-lived effects increases with the number of laser pulses, the overall pump-induced change becomes increasingly dominated by these long-lived contributions as more laser pulses per train are included. Hence, one approach to limiting these effects is to reduce the number of laser pulses per train, which comes at the detriment of data quality or measurement time. We have previously shown that by limiting the number of laser pulses we use for the data analysis and additionally compensating for the evaluation of uncertainty while modelling, we can deal successfully with these modifications \cite{lojewski_interplay_2023,lojewski_photo-induced_2024}. In this paper, we want to present a different approach where we aim to separate and analyze the contribution of these long-lasting effects individually from the pump-induced modifications, as well as attempt to correct the $\Delta$XAS measured with a higher number of pulses. To that end, we applied our modelling approach to further evaluate their shape.

            \subsection{Modelling}
            
                \begin{figure}
                    \caption{(a) Modelling of $\Delta$XAS for 1 and 16 laser pulses of the O K, NiO L and Ni L-edges, with the corresponding modelled heating contribution shown as a dashed line. (b) Comparision of average measured and modelled heating contribution after 4,8,12 and 16 for all pump fluences. (c) Modelling parameter for the energy shift $\Delta$E and broadening $\Delta \omega$ for the two fluences as a function of accumulated laser pulses.}
                    \includegraphics[width=1\textwidth]{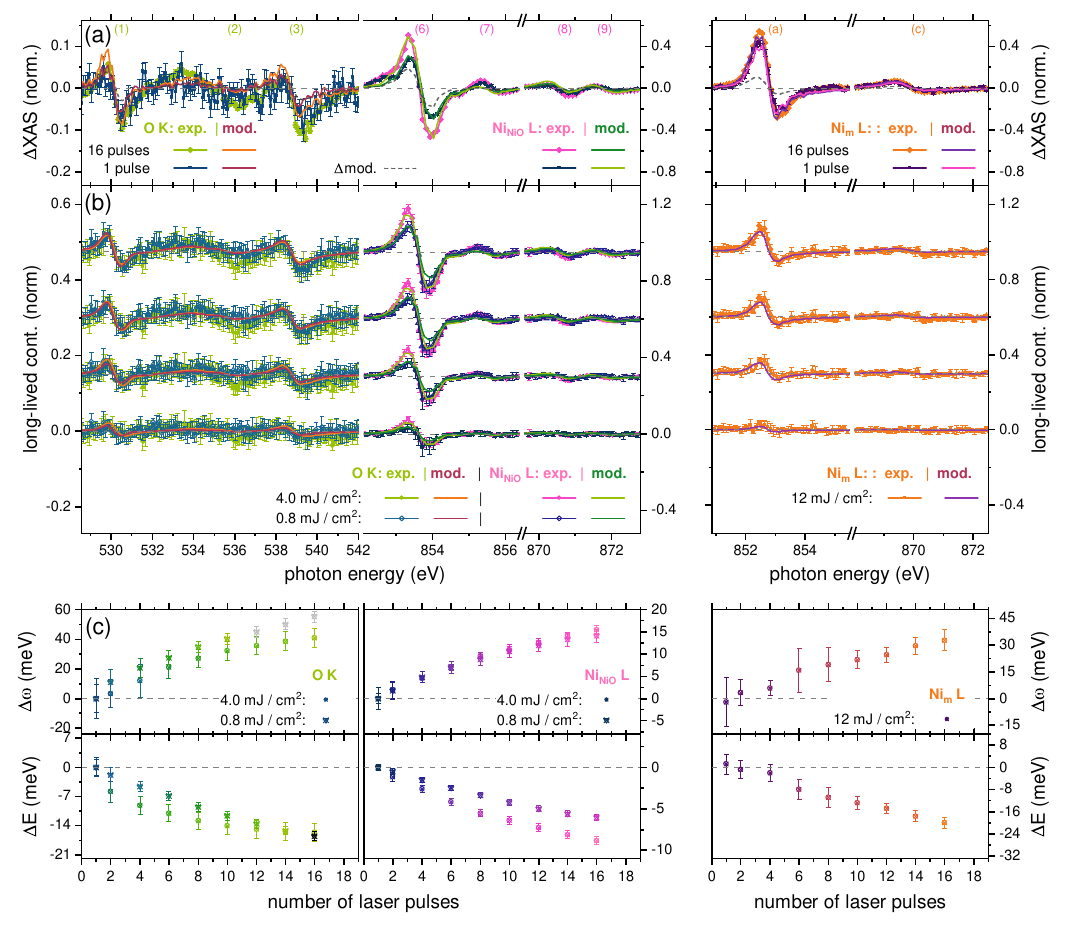}
                    \label{fig_7}
                \end{figure}
            
            We model the contribution of the long-lived effects by modelling the $\Delta$XAS after multiple laser pulses per train and comparing how the modelled changes compared to the modelled $\Delta$XAS after one laser pulse. To that end, we model, as earlier described, the $\Delta$XAS after one, two and every subsequent two laser pulses simultaneously. As the groundstate spectrum remains unmodified regardless of the number of X-ray pulses per train except for the level of noise, we use the spectrum measured with the maximum number of X-ray pulses as a basis to model all $\Delta$XAS signals in order to obtain the best modelling quality. Modelling all curves simultaneously allows us to also set one parameter, either the energy shift $\Delta$E or broadening $\Delta \omega$ to be optimized globally for all $\Delta$XAS signals, using only the remaining free parameter to model the contribution of the long-lived excitations. However, overall, we found that to obtain the best result, both parameters should be varied to describe the contribution of the long-lived excitations for all measurements across all absorption edges. We obtain the contribution of the long-lived excitations as before by calculating the difference between the modelled $\Delta$XAS after multiple laser pulses per train and the modelled first initial laser excitation. This approach is shown exemplarily in figure~\ref{fig_7} panel (a) for a measurement with a time delay $\Delta$t of 0.5~ps for the NiO absorption edges and 0.4~ps for the Ni-metal L-edges. We average the different modelled contributions to obtain an average modelled contribution of the long-lived excitations, which is shown in comparison with the average measured contribution in figure~\ref{fig_7} panel (b).
            
            Overall, the modelling manages to capture the shape of the contribution fairly well for all absorption edges, with only some minor deviations. At the oxygen K-edge, the modifications of the second fine structure peak (2) are not captured well for a higher number of excitations per train. The contribution of long-lived excitations around the first and third fine structure peaks (1,3) seems mostly well described, although the modelling itself does not work too well in these regions for the $\Delta$XAS. For the NiO L$_3$-edge, the model works extremely well in describing both the overall $\Delta$XAS as well as the contribution of long-lived excitations. The only deviations that are visible are at the pre-edge feature, which results from non-equilibrium multiple excitations, is unaffected by the long-lived effects and also not captured by the modelling \cite{lojewski_photo-induced_2024}, as well as the L$_3$ satellite (7) where both the modelling of the $\Delta$XAS and the long-lived contributions fails. Additionally, there are also differences visible in both the modelling of the $\Delta$XAS and the long-lived contributions at the first peak of the NiO L$_2$-edge (8). At the Ni-metal L-edges, the modelling also works fairly well, accurately capturing the shape of both the overall $\Delta$XAS as well as the contribution of the long-lived excitations, with the only deviation being visible at the reversal point between the positive and negative induced change (852.8~eV). Panel (c) of figure~\ref{fig_7} shows the change of the determined energy shift $\Delta$E and broadening $\Delta \omega$ for increasing the number of laser pulses per train. 
            
            For the Ni metal L-edges, we find that both the energy shift and broadening increase linearly, with the number of laser pulses per train. At the NiO L-edges, the shift and broadening both seem to saturate. However, while the energy shift shows a strong fluence dependence, we find no influence of the fluence on the determined broadening. This difference in the energy shift can be directly observed in the shape of the long-lived contribution as the positive contribution around 853.3~eV increases faster with the number of pulses per train, while the negative contribution at 853.8~eV stays mostly the same.
            
            For the O K-edge the energy and broadening also show a saturating type behaviour. However, we find the behaviour of the fluence dependence to be directly contradictory to the changes in the absolute area observed earlier. We find that the value of the broadening seems to increase faster with the lower pump fluence. The determined energy shift, although initially increasing faster, also seems to saturate earlier, while the energy shift for 0.8~mJ/cm$^2$ continues to increase further. This contradiction is most likely a combination of the higher noise level at the O K-edge and the overall model not completely accurately describing the $\Delta$XAS. We previously emphasized the modification around the second fine-structure peak (2) as the most prominent difference between 0.8 and 4.0~mJ/cm$^2$, which is incidentally not accurately captured by the modelling. From this, we conclude that the modelling approach is not well suited to describe the differences between the two pump fluences at the O K-edge. 
        
        \section{Discussion}
        
            We have previously discussed the effects of long-lived excitations in Ni-metal as a local heat increase \cite{le_guyader_photon-shot-noise-limited_2023}. The local heating is the consequence of the sample geometry needed for the employed setup, requiring the Ni-layer to be sandwiched between an X-ray transparent but also weakly heat-conducting substrate and the insulating capping layer to prevent oxidation. Thus, the heat transport out of plane is severely limited, and heat transport can only occur in-plane. As a result, the energy is not completely dissipated between the laser pulses during a train, so the heat of the pumped spot continuously increases within a train. We have now extracted the secondary modification that results from this local heat increase and found that it mirrors the shape of the initial induced changes but with reduced intensity. Exactly this behaviour has been previously observed by T. Kachel et al. in Ni-metal for laser-induced heating after 100~ps, where they identified an energy shift of 38~meV  \cite{kachel_transient_2009}. We find comparable values of the energy shift, which reaches 20$\pm$2~meV after 16 laser pulses. This further cements this modification as a local increase in the temperature, as we previously outlined \cite{le_guyader_photon-shot-noise-limited_2023}.
            
            For NiO, the contribution of the long-lived excitations increases non-linearly, which we would not expect for a purely local heating-induced effect. For metal oxides, it was previously shown that charge carrier trapping through exciton \cite{foglia_revealing_2019} and polaron \cite{biswas_ultrafast_2018,carneiro_excitation-wavelength-dependent_2017} formation causes long-lived excitations. These excitations have been shown to be extremely long-lived, existing for hundreds of ns \cite{foglia_revealing_2019,santomauro_localized_2017}, making them a potential reason for different behaviour observed in NiO. Theoretical calculations show that for NiO, small polaron formation is possible by trapping a photoexcited electron on a Ni-site through an extension of the surrounding oxygen atoms \cite{yu_charge_2012}. The formation of polarons in NiO has been previously observed by S. Biswas et al. \cite{biswas_ultrafast_2018} in XUV spectroscopy of the Ni M$_{2,3}$-edge as a reduction of photo-induced redshift, with the reduced redshift persisting over tens of ps. This could link the occurrence of the long-lived energy shift that we observe at the Ni L-edges of NiO to the formation of polarons. Additionally, in black $\gamma$-CsPbI$_3$ it was demonstrated that the formation of polarons not only scales with the laser fluence but also reaches a stable regime where the polaron formation saturates \cite{zhang_stable_2023}. This could further explain the saturating behaviour of the energy shift that we observe at the NiO Ni L-edges. In addition, the formation of polarons could also be related to the long-lived modification that we observe at the O K-edge. As the polarons are stabilized through an extension of the oxygen atoms surrounding the Ni site, this should influence the Oxygen K-edge spectrum, especially as it was shown that it is sensitive to lattice excitations in MgO \cite{rothenbach_microscopic_2019,rothenbach_effect_2021}.

            \subsection{Correction Methodes}
        
                \begin{figure}
                    \caption{Comparing the $\Delta$XAS after one and sixteen accumulated laser pulses as measured, using the modelled long-lived contribution for correction and using the average long-lived contribution for correction for the Ni-metal L-edges (a), NiO L-edges (b) and O K-edge (c).}
                    \includegraphics[width=8.85cm]{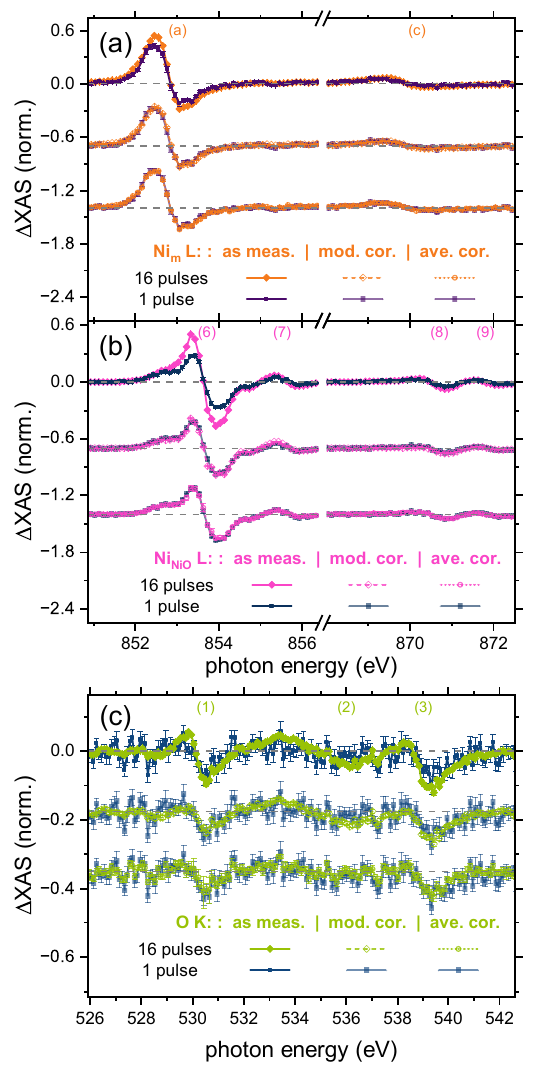}
                    \label{fig_8}
                \end{figure}
        
            We have previously discussed that the effects of these long-lived excitations can be minimized by limiting the number of laser pulses per train, and while this is a valid approach, it might not always be the best option. If the pump-induced change is relatively small compared to the noise level, it might not be possible to limit the number of pulses per train too much, and this, in combination with the non-linear behaviour observed in NiO, can lead to a significant contribution in the $\Delta$XAS. With our analysis, we could separate the contribution of the long-lived excitations from the initial laser-induced excitations and describe their main underlying modifications using our modelling approach. Now, we want to apply these results to discuss two possible correction methods that could be used to disentangle the initial pump-induced changes from the long-lived effects while utilizing more pulses per train. For one, we can use our modelled contribution of the long-lived excitations to correct the measured $\Delta$XAS. Alternatively, the average heating contribution can also be used to apply a correction. Figure~\ref{fig_8} shows a comparison between the $\Delta$XAS measured with one or sixteen pulses per train for the Ni-metal L-edges (a), NiO L-edges (b) and O K-edge (c) and the results of the two correction methods. In general, using the modelled contribution for the correction gives the best signal-to-noise ratio but has a big disadvantage in that it only works if the modelling completely accurately captures all long-lived modifications. This is not the case with our simplified modelling approach, which leads to some visible deviations at the NiO L$_3$ Satellite (7) and the second fine structure peak of the O K-edge (2). Alternatively, the average measured contribution can also be used to correct the $\Delta$XAS, which is shown at the bottom of panel (a-c). This approach gives the best result for the overall correction but shows a signal-to-noise ratio that is a bit worse than the correction using the modelled contribution. The advantage of using the average over the model is that the more scans that have been measured, the better the correction will work. In general, both these correction methods appear to work relatively well, but overall, we would still recommend reducing the number of laser pulses per train over these correction methods if applicable or adjusting the measurement pattern to reduce the effects. However, if the contribution of the long-lived states increases rapidly with the first few laser pulses and the pump-induced change is relatively small compared to these contributions, as we showed here with the O K-edge, one of the two proposed corrections may be used.  
        
        \section{Conclusion}
        
            We showed the effects of long-lived excitations in high-repetition rate stroboscopic transient X-ray absorption spectroscopy measurements on nickel and nickel oxide. We identified these long-lived excitations as an additional modification of the $\Delta$XAS that appears and increases with an increasing number of laser pulses during the short burst measurements (trains) of the European X-ray free electron laser. In the here presented analysis, we could disentangle their contribution from the initial pump-induced change and analyze them separately using our modelling approach. In Ni, the long-lived effects lead to a linearly increasing broadening and energy shift of the Ni L-edges. The linear behaviour and shape of these contributions links the effect to local laser-induced heating, which is the result of the sample geometry limiting the only efficient heat transport to be in-plane. In NiO, these effects lead to increased broadening and energy shift of the Ni L-edges, which appear to saturate for a higher number of pulses. Additionally, the O K-edge also shows an increased broadening and energy shift, which indicates a saturating behaviour. By disentangling these long-lived contributions, we found that the energy shift of the Ni L-edges is strongly affected by the excitation fluence while the broadening is mostly unaffected. Long-lived energy shifts have been previously observed in NiO and were linked to charge carrier trapping through polaron formation on a Nickel atom site, which is stabilized through an oxygen lattice expansion. Polaron formation could also explain the strong dependence of the energy shift on the excitation fluence and the saturating behaviour of these long-lived excitations. Furthermore, the expansion of the oxygen lattice due to the polarion formation could be linked to the modification of the O K-edge. Finally, we used our evaluation and modelling of these contributions to devise two correction methods that can help extract the initial pump-induced change without limiting the number of laser pulses per train, which allows for a better signal-to-noise ratio.

        \section{Data availability}\label{Data_availability}
        
            The raw data on Ni and NiO shown in this publication are publicly available. The raw data measured on the Ni samples in experiment UP2161 is available at\\ 
            https://doi.org/10.22003/XFEL.EU-DATA-002161-00 \cite{noauthor_Ni_Data_2020}. The raw data measured on the NiO samples in experiment UP2589 is available at\\ https://doi.org/10.22003/XFEL.EU-DATA-002589-00 \cite{noauthor_NiO_Data_2021}. \\


        \ack{Acknowledgements}
        
            We acknowledge the European XFEL in Schenefeld, Germany, for the provision of X-ray free-electron laser beamtimes UP2161 \cite{noauthor_Ni_Data_2020} and UP2589 \cite{noauthor_NiO_Data_2021} at the SCS instrument and thank the staff for their assistance. We thank Prof.~Dr.~Peter~Fischer, Dr.~Charles-Henri~Lambert and Dr.~Jun~Zhu for their support and expertise during experiment UP2161. We thank Prof.~Dr.~Gheorghe~S.~Chiuzb\u{a}ian for his contribution to the experiment UP2589. Funded by the Deutsche Forschungsgemeinschaft (DFG, German Research Foundation) through  Project No. 278162697 - SFB 1242. U.B. acknowledges support from the Deutsche Forschungsgemeinschaft (DFG, German Science Foundation) through FOR 5249-449872909 (Project P6). P.~S.~M., R.~Y.~E. and M.~B. acknowledge funding from the Helmholtz Association [grant number VH-NG-1105].
    

        \referencelist{iucr}

    

@article{ackermann_operation_2007,
	title = {Operation of a free-electron laser from the extreme ultraviolet to the water window},
	volume = {1},
	issn = {1749-4885 1749-4893},
	doi = {10.1038/nphoton.2007.76},
	number = {6},
	journal = {Nature Photonics},
	author = {Ackermann, W. and Asova, G. and Ayvazyan, V. and Azima, A. and Baboi, N. and Bähr, J. and Balandin, V. and Beutner, B. and Brandt, A. and Bolzmann, A. and Brinkmann, R. and Brovko, O. I. and Castellano, M. and Castro, P. and Catani, L. and Chiadroni, E. and Choroba, S. and Cianchi, A. and Costello, J. T. and Cubaynes, D. and Dardis, J. and Decking, W. and Delsim-Hashemi, H. and Delserieys, A. and Di Pirro, G. and Dohlus, M. and Düsterer, S. and Eckhardt, A. and Edwards, H. T. and Faatz, B. and Feldhaus, J. and Flöttmann, K. and Frisch, J. and Fröhlich, L. and Garvey, T. and Gensch, U. and Gerth, Ch and Görler, M. and Golubeva, N. and Grabosch, H. J. and Grecki, M. and Grimm, O. and Hacker, K. and Hahn, U. and Han, J. H. and Honkavaara, K. and Hott, T. and Hüning, M. and Ivanisenko, Y. and Jaeschke, E. and Jalmuzna, W. and Jezynski, T. and Kammering, R. and Katalev, V. and Kavanagh, K. and Kennedy, E. T. and Khodyachykh, S. and Klose, K. and Kocharyan, V. and Körfer, M. and Kollewe, M. and Koprek, W. and Korepanov, S. and Kostin, D. and Krassilnikov, M. and Kube, G. and Kuhlmann, M. and Lewis, C. L. S. and Lilje, L. and Limberg, T. and Lipka, D. and Löhl, F. and Luna, H. and Luong, M. and Martins, M. and Meyer, M. and Michelato, P. and Miltchev, V. and Möller, W. D. and Monaco, L. and Müller, W. F. O. and Napieralski, O. and Napoly, O. and Nicolosi, P. and Nölle, D. and Nuñez, T. and Oppelt, A. and Pagani, C. and Paparella, R. and Pchalek, N. and Pedregosa-Gutierrez, J. and Petersen, B. and Petrosyan, B. and Petrosyan, G. and Petrosyan, L. and Pflüger, J. and Plönjes, E. and Poletto, L. and Pozniak, K. and Prat, E. and Proch, D. and Pucyk, P. and Radcliffe, P. and Redlin, H. and Rehlich, K. and Richter, M. and Roehrs, M. and Roensch, J. and Romaniuk, R. and Ross, M. and Rossbach, J. and Rybnikov, V. and Sachwitz, M. and Saldin, E. L. and Sandner, W. and Schlarb, H. and Schmidt, B. and Schmitz, M. and Schmüser, P. and Schneider, J. R. and Schneidmiller, E. A. and Schnepp, S. and Schreiber, S. and Seidel, M. and Sertore, D. and Shabunov, A. V. and Simon, C. and Simrock, S. and Sombrowski, E. and Sorokin, A. A. and Spanknebel, P. and Spesyvtsev, R. and Staykov, L. and Steffen, B. and Stephan, F. and Stulle, F. and Thom, H. and Tiedtke, K. and Tischer, M. and Toleikis, S. and Treusch, R. and Trines, D. and Tsakov, I. and Vogel, E. and Weiland, T. and Weise, H. and Wellhöfer, M. and Wendt, M. and Will, I. and Winter, A. and Wittenburg, K. and Wurth, W. and Yeates, P. and Yurkov, M. V. and Zagorodnov, I. and Zapfe, K.},
	year = {2007},
	pages = {336--342},
}

@inproceedings{allaria_flash2020_2021,
	title = {{FLASH2020}+ {Plans} for a {New} {Coherent} {Source} at {DESY}},
	url = {https://bib-pubdb1.desy.de/record/472675},
	doi = {10.18429/JACOW-IPAC2021-TUPAB086},
	booktitle = {Proceedings of the 12th {International} {Particle} {Accelerator} {Conference}, {IPAC2021}, {Campinas}, {SP}, {Brazil}},
	publisher = {JACoW Publishing, Geneva, Switzerland},
	author = {Allaria, Enrico and Baboi, Nicoleta and Baev, Karolin and Beye, Martin and Brenner, Günter and Christie, Florian and Gerth, Christopher and Hartl, Ingmar and Honkavaara, Katja and Manschwetus, Bastian and Mueller-Dieckmann, Jochen and Pan, Rui and Plönjes-Palm, Elke and Rasmussen, Olaf and Rönsch-Schulenburg, Juliane and Schaper, Lucas and Schneidmiller, Evgeny and Schreiber, Siegfried and Tiedtke, Kai and Tischer, Markus and Toleikis, Sven and Treusch, Rolf and Vogt, Mathias and Winkelmann, Lutz and Yurkov, Mikhail and Zemella, Johann},
	month = may,
	year = {2021},
	note = {Backup Publisher: 12th International Particle Accelerator Conference, Campinas (Brazil), 24 May 2021 - 28 May 2021},
	pages = {4},
}

@article{biswas_ultrafast_2018,
	title = {Ultrafast {Electron} {Trapping} and {Defect}-{Mediated} {Recombination} in {NiO} {Probed} by {Femtosecond} {Extreme} {Ultraviolet} {Reflection}–{Absorption} {Spectroscopy}},
	volume = {9},
	url = {https://doi.org/10.1021/acs.jpclett.8b01865},
	doi = {10.1021/acs.jpclett.8b01865},
	number = {17},
	urldate = {2024-08-23},
	journal = {The Journal of Physical Chemistry Letters},
	author = {Biswas, Somnath and Husek, Jakub and Londo, Stephen and Baker, L. Robert},
	month = sep,
	year = {2018},
	note = {Publisher: American Chemical Society},
	pages = {5047--5054},
}

@inproceedings{brachmann_lcls-ii_2019,
	series = {Free {Electron} {Laser} {Conference}},
	title = {{LCLS}-{II} - {Status} and {Upgrades}},
	isbn = {978-3-95450-210-3},
	url = {http://jacow.org/fel2019/papers/fra02.pdf},
	doi = {10.18429/JACoW-FEL2019-FRA02},
	language = {english},
	booktitle = {Proc. {FEL}'19},
	publisher = {JACoW Publishing, Geneva, Switzerland},
	author = {Brachmann, A. and Dunham, M. and Schmerge, J. F.},
	month = nov,
	year = {2019},
	note = {ISSN: ""
Issue: 39},
	pages = {772--775},
}

@article{carneiro_excitation-wavelength-dependent_2017,
	title = {Excitation-wavelength-dependent small polaron trapping of photoexcited carriers in alpha-{Fe}(2){O}(3)},
	volume = {16},
	issn = {1476-4660 (Electronic) 1476-1122 (Linking)},
	doi = {10.1038/nmat4936},
	number = {8},
	journal = {Nat Mater},
	author = {Carneiro, L. M. and Cushing, S. K. and Liu, C. and Su, Y. and Yang, P. and Alivisatos, A. P. and Leone, S. R.},
	month = aug,
	year = {2017},
	pages = {819--825},
}

@article{carva_influence_2009,
	title = {Influence of laser-excited electron distributions on the {X}-ray magnetic circular dichroism spectra: {Implications} for femtosecond demagnetization in {Ni}},
	volume = {86},
	issn = {0295-5075},
	shorttitle = {Influence of laser-excited electron distributions on the {X}-ray magnetic circular dichroism spectra},
	url = {https://dx.doi.org/10.1209/0295-5075/86/57002},
	doi = {10.1209/0295-5075/86/57002},
	language = {en},
	number = {5},
	urldate = {2024-10-31},
	journal = {Europhysics Letters},
	author = {Carva, K. and Legut, D. and Oppeneer, P. M.},
	month = jun,
	year = {2009},
	pages = {57002},
}

@article{decking_mhz-repetition-rate_2020,
	title = {A {MHz}-repetition-rate hard {X}-ray free-electron laser driven by a superconducting linear accelerator},
	volume = {14},
	issn = {1749-4885 1749-4893},
	doi = {10.1038/s41566-020-0607-z},
	number = {6},
	journal = {Nature Photonics},
	author = {Decking, W. and Abeghyan, S. and Abramian, P. and Abramsky, A. and Aguirre, A. and Albrecht, C. and Alou, P. and Altarelli, M. and Altmann, P. and Amyan, K. and Anashin, V. and Apostolov, E. and Appel, K. and Auguste, D. and Ayvazyan, V. and Baark, S. and Babies, F. and Baboi, N. and Bak, P. and Balandin, V. and Baldinger, R. and Baranasic, B. and Barbanotti, S. and Belikov, O. and Belokurov, V. and Belova, L. and Belyakov, V. and Berry, S. and Bertucci, M. and Beutner, B. and Block, A. and Blöcher, M. and Böckmann, T. and Bohm, C. and Böhnert, M. and Bondar, V. and Bondarchuk, E. and Bonezzi, M. and Borowiec, P. and Bösch, C. and Bösenberg, U. and Bosotti, A. and Böspflug, R. and Bousonville, M. and Boyd, E. and Bozhko, Y. and Brand, A. and Branlard, J. and Briechle, S. and Brinker, F. and Brinker, S. and Brinkmann, R. and Brockhauser, S. and Brovko, O. and Brück, H. and Brüdgam, A. and Butkowski, L. and Büttner, T. and Calero, J. and Castro-Carballo, E. and Cattalanotto, G. and Charrier, J. and Chen, J. and Cherepenko, A. and Cheskidov, V. and Chiodini, M. and Chong, A. and Choroba, S. and Chorowski, M. and Churanov, D. and Cichalewski, W. and Clausen, M. and Clement, W. and Cloué, C. and Cobos, J. A. and Coppola, N. and Cunis, S. and Czuba, K. and Czwalinna, M. and D’Almagne, B. and Dammann, J. and Danared, H. and de Zubiaurre Wagner, A. and Delfs, A. and Delfs, T. and Dietrich, F. and Dietrich, T. and Dohlus, M. and Dommach, M. and Donat, A. and Dong, X. and Doynikov, N. and Dressel, M. and Duda, M. and Duda, P. and Eckoldt, H. and Ehsan, W. and Eidam, J. and Eints, F. and Engling, C. and Englisch, U. and Ermakov, A. and Escherich, K. and Eschke, J. and Saldin, E. and Faesing, M. and Fallou, A. and Felber, M. and Fenner, M. and Fernandes, B. and Fernández, J. M. and Feuker, S. and Filippakopoulos, K. and Floettmann, K. and Fogel, V. and Fontaine, M. and Francés, A. and Martin, I. Freijo and Freund, W. and Freyermuth, T. and Friedland, M. and Fröhlich, L. and Fusetti, M. and Fydrych, J. and Gallas, A. and García, O. and Garcia-Tabares, L. and Geloni, G. and Gerasimova, N. and Gerth, C. and Geßler, P. and Gharibyan, V. and Gloor, M. and Głowinkowski, J. and Goessel, A. and Gołębiewski, Z. and Golubeva, N. and Grabowski, W. and Graeff, W. and Grebentsov, A. and Grecki, M. and Grevsmuehl, T. and Gross, M. and Grosse-Wortmann, U. and Grünert, J. and Grunewald, S. and Grzegory, P. and Feng, G. and Guler, H. and Gusev, G. and Gutierrez, J. L. and Hagge, L. and Hamberg, M. and Hanneken, R. and Harms, E. and Hartl, I. and Hauberg, A. and Hauf, S. and Hauschildt, J. and Hauser, J. and Havlicek, J. and Hedqvist, A. and Heidbrook, N. and Hellberg, F. and Henning, D. and Hensler, O. and Hermann, T. and Hidvégi, A. and Hierholzer, M. and Hintz, H. and Hoffmann, F. and Hoffmann, Markus and Hoffmann, Matthias and Holler, Y. and Hüning, M. and Ignatenko, A. and Ilchen, M. and Iluk, A. and Iversen, J. and Iversen, J. and Izquierdo, M. and Jachmann, L. and Jardon, N. and Jastrow, U. and Jensch, K. and Jensen, J. and Jeżabek, M. and Jidda, M. and Jin, H. and Johansson, N. and Jonas, R. and Kaabi, W. and Kaefer, D. and Kammering, R. and Kapitza, H. and Karabekyan, S. and Karstensen, S. and Kasprzak, K. and Katalev, V. and Keese, D. and Keil, B. and Kholopov, M. and Killenberger, M. and Kitaev, B. and Klimchenko, Y. and Klos, R. and Knebel, L. and Koch, A. and Koepke, M. and Köhler, S. and Köhler, W. and Kohlstrunk, N. and Konopkova, Z. and Konstantinov, A. and Kook, W. and Koprek, W. and Körfer, M. and Korth, O. and Kosarev, A. and Kosiński, K. and Kostin, D. and Kot, Y. and Kotarba, A. and Kozak, T. and Kozak, V. and Kramert, R. and Krasilnikov, M. and Krasnov, A. and Krause, B. and Kravchuk, L. and Krebs, O. and Kretschmer, R. and Kreutzkamp, J. and Kröplin, O. and Krzysik, K. and Kube, G. and Kuehn, H. and Kujala, N. and Kulikov, V. and Kuzminych, V. and La Civita, D. and Lacroix, M. and Lamb, T. and Lancetov, A. and Larsson, M. and Le Pinvidic, D. and Lederer, S. and Lensch, T. and Lenz, D. and Leuschner, A. and Levenhagen, F. and Li, Y. and Liebing, J. and Lilje, L. and Limberg, T. and Lipka, D. and List, B. and Liu, J. and Liu, S. and Lorbeer, B. and Lorkiewicz, J. and Lu, H. H. and Ludwig, F. and Machau, K. and Maciocha, W. and Madec, C. and Magueur, C. and Maiano, C. and Maksimova, I. and Malcher, K. and Maltezopoulos, T. and Mamoshkina, E. and Manschwetus, B. and Marcellini, F. and Marinkovic, G. and Martinez, T. and Martirosyan, H. and Maschmann, W. and Maslov, M. and Matheisen, A. and Mavric, U. and Meißner, J. and Meissner, K. and Messerschmidt, M. and Meyners, N. and Michalski, G. and Michelato, P. and Mildner, N. and Moe, M. and Moglia, F. and Mohr, C. and Mohr, S. and Möller, W. and Mommerz, M. and Monaco, L. and Montiel, C. and Moretti, M. and Morozov, I. and Morozov, P. and Mross, D. and Mueller, J. and Müller, C. and Müller, J. and Müller, K. and Munilla, J. and Münnich, A. and Muratov, V. and Napoly, O. and Näser, B. and Nefedov, N. and Neumann, Reinhard and Neumann, Rudolf and Ngada, N. and Noelle, D. and Obier, F. and Okunev, I. and Oliver, J. A. and Omet, M. and Oppelt, A. and Ottmar, A. and Oublaid, M. and Pagani, C. and Paparella, R. and Paramonov, V. and Peitzmann, C. and Penning, J. and Perus, A. and Peters, F. and Petersen, B. and Petrov, A. and Petrov, I. and Pfeiffer, S. and Pflüger, J. and Philipp, S. and Pienaud, Y. and Pierini, P. and Pivovarov, S. and Planas, M. and Pławski, E. and Pohl, M. and Polinski, J. and Popov, V. and Prat, S. and Prenting, J. and Priebe, G. and Pryschelski, H. and Przygoda, K. and Pyata, E. and Racky, B. and Rathjen, A. and Ratuschni, W. and Regnaud-Campderros, S. and Rehlich, K. and Reschke, D. and Robson, C. and Roever, J. and Roggli, M. and Rothenburg, J. and Rusiński, E. and Rybaniec, R. and Sahling, H. and Salmani, M. and Samoylova, L. and Sanzone, D. and Saretzki, F. and Sawlanski, O. and Schaffran, J. and Schlarb, H. and Schlösser, M. and Schlott, V. and Schmidt, C. and Schmidt-Foehre, F. and Schmitz, M. and Schmökel, M. and Schnautz, T. and Schneidmiller, E. and Scholz, M. and Schöneburg, B. and Schultze, J. and Schulz, C. and Schwarz, A. and Sekutowicz, J. and Sellmann, D. and Semenov, E. and Serkez, S. and Sertore, D. and Shehzad, N. and Shemarykin, P. and Shi, L. and Sienkiewicz, M. and Sikora, D. and Sikorski, M. and Silenzi, A. and Simon, C. and Singer, W. and Singer, X. and Sinn, H. and Sinram, K. and Skvorodnev, N. and Smirnow, P. and Sommer, T. and Sorokin, A. and Stadler, M. and Steckel, M. and Steffen, B. and Steinhau-Kühl, N. and Stephan, F. and Stodulski, M. and Stolper, M. and Sulimov, A. and Susen, R. and Świerblewski, J. and Sydlo, C. and Syresin, E. and Sytchev, V. and Szuba, J. and Tesch, N. and Thie, J. and Thiebault, A. and Tiedtke, K. and Tischhauser, D. and Tolkiehn, J. and Tomin, S. and Tonisch, F. and Toral, F. and Torbin, I. and Trapp, A. and Treyer, D. and Trowitzsch, G. and Trublet, T. and Tschentscher, T. and Ullrich, F. and Vannoni, M. and Varela, P. and Varghese, G. and Vashchenko, G. and Vasic, M. and Vazquez-Velez, C. and Verguet, A. and Vilcins-Czvitkovits, S. and Villanueva, R. and Visentin, B. and Viti, M. and Vogel, E. and Volobuev, E. and Wagner, R. and Walker, N. and Wamsat, T. and Weddig, H. and Weichert, G. and Weise, H. and Wenndorf, R. and Werner, M. and Wichmann, R. and Wiebers, C. and Wiencek, M. and Wilksen, T. and Will, I. and Winkelmann, L. and Winkowski, M. and Wittenburg, K. and Witzig, A. and Wlk, P. and Wohlenberg, T. and Wojciechowski, M. and Wolff-Fabris, F. and Wrochna, G. and Wrona, K. and Yakopov, M. and Yang, B. and Yang, F. and Yurkov, M. and Zagorodnov, I. and Zalden, P. and Zavadtsev, A. and Zavadtsev, D. and Zhirnov, A. and Zhukov, A. and Ziemann, V. and Zolotov, A. and Zolotukhina, N. and Zummack, F. and Zybin, D.},
	year = {2020},
	pages = {391--397},
}

@article{doring_multifocus_2020,
	title = {Multifocus off-axis zone plates for x-ray free-electron laser experiments},
	volume = {7},
	copyright = {© 2020 Optical Society of America},
	issn = {2334-2536},
	url = {https://opg.optica.org/optica/abstract.cfm?uri=optica-7-8-1007},
	doi = {10.1364/OPTICA.398022},
	language = {EN},
	number = {8},
	urldate = {2024-11-27},
	journal = {Optica},
	author = {Döring, Florian and Rösner, Benedikt and Langer, Manuel and Kubec, Adam and Kleibert, Armin and Raabe, Jörg and Vaz, Carlos A. F. and Lebugle, Maxime and David, Christian},
	month = aug,
	year = {2020},
	note = {Publisher: Optica Publishing Group},
	pages = {1007--1014},
}

@article{emma_first_2010,
	title = {First lasing and operation of an ångstrom-wavelength free-electron laser},
	volume = {4},
	issn = {1749-4885 1749-4893},
	doi = {10.1038/nphoton.2010.176},
	number = {9},
	journal = {Nature Photonics},
	author = {Emma, P. and Akre, R. and Arthur, J. and Bionta, R. and Bostedt, C. and Bozek, J. and Brachmann, A. and Bucksbaum, P. and Coffee, R. and Decker, F. J. and Ding, Y. and Dowell, D. and Edstrom, S. and Fisher, A. and Frisch, J. and Gilevich, S. and Hastings, J. and Hays, G. and Hering, Ph and Huang, Z. and Iverson, R. and Loos, H. and Messerschmidt, M. and Miahnahri, A. and Moeller, S. and Nuhn, H. D. and Pile, G. and Ratner, D. and Rzepiela, J. and Schultz, D. and Smith, T. and Stefan, P. and Tompkins, H. and Turner, J. and Welch, J. and White, W. and Wu, J. and Yocky, G. and Galayda, J.},
	year = {2010},
	pages = {641--647},
}

@article{eschenlohr_ultrafast_2013,
	title = {Ultrafast spin transport as key to femtosecond demagnetization},
	volume = {12},
	issn = {1476-4660 (Electronic) 1476-1122 (Linking)},
	doi = {10.1038/nmat3546},
	number = {4},
	journal = {Nat Mater},
	author = {Eschenlohr, A. and Battiato, M. and Maldonado, P. and Pontius, N. and Kachel, T. and Holldack, K. and Mitzner, R. and Fohlisch, A. and Oppeneer, P. M. and Stamm, C.},
	month = apr,
	year = {2013},
	pages = {332--6},
}

@article{eschenlohr_spin_2017,
	title = {Spin currents during ultrafast demagnetization of ferromagnetic bilayers},
	volume = {29},
	issn = {1361-648X (Electronic) 0953-8984 (Linking)},
	doi = {10.1088/1361-648X/aa7dd3},
	number = {38},
	journal = {J Phys Condens Matter},
	author = {Eschenlohr, A. and Persichetti, L. and Kachel, T. and Gabureac, M. and Gambardella, P. and Stamm, C.},
	month = sep,
	year = {2017},
	pages = {384002},
}

@article{eschenlohr_spin_2021,
	title = {Spin dynamics at interfaces on femtosecond timescales},
	volume = {33},
	issn = {1361-648X (Electronic) 0953-8984 (Linking)},
	doi = {10.1088/1361-648X/abb519},
	number = {1},
	journal = {J Phys Condens Matter},
	author = {Eschenlohr, A.},
	month = jan,
	year = {2021},
	pages = {013001},
}

@inproceedings{fangohr_data_2020,
	title = {Data {Exploration} and {Analysis} with {Jupyter} {Notebooks}},
	isbn = {978-3-95450-209-7},
	url = {https://accelconf.web.cern.ch/icalepcs2019/doi/JACoW-ICALEPCS2019-TUCPR02.html},
	doi = {10.18429/JACoW-ICALEPCS2019-TUCPR02},
	language = {en},
	urldate = {2024-08-26},
	publisher = {JACOW Publishing, Geneva, Switzerland},
	author = {Fangohr, Hans and Beg, Marijan and Bergemann, Martin and Bondar, Valerii and Brockhauser, Sandor and Campbell, Aidan and Carinan, Cammille and Costa, Raul and Dall'Antonia, Fabio and Danilevski, Cyril and E, Juncheng and Ehsan, Wajid and Esenov, Sergey and Fabbri, Riccardo and Fangohr, Susanne and Fernandez-del-Castillo, Enol and Flucke, Gero and Fortmann-Grote, Carsten and Fulla Marsa, Daniel and Giovanetti, Gabriele and Goeries, Dennis and Götz, Andrew and Hall, Jamie and Hauf, Steffen and Hickin, David and Holm Rod, Thomas and Jarosiewicz, Tobiasz and Kamil, Ebad and Karnevskiy, Mikhail and Kieffer, Jerome and Kirienko, Yury and Klimovskaia, Anna and Kluyver, Thomas and Kuster, Markus and Le Guyader, Loïc and Madsen, Anders and Maia, Luis and Mamchyk, Denys and Mercadier, Laurent and Michelat, Thomas and Möller, Johannes and Mohacsi, Istvan and Parenti, Andrea and Pellegrini, Eric and Perrin, Jean-Francois and Reiser, Mario and Reppin, Johannes and Rosca, Robert and Rück, Denivy and Rüter, Tonn and Santos, Hugo and Schaffer, Robert and Scherz, Andreas and Schlünzen, Frank and Scholz, Markus and Schuh, Michael and Selknaes, Jesper Rude and Silenzi, Alessandro and Sipos, Gergely and Spirzewski, Michal and Sztuk, Jolanta and Szuba, Janusz and Taylor, Jonathan and Trojanowski, Sebastian and Wrona, Krzysztof and Yaroslavtsev, Alexander and Zhu, Jun},
	month = aug,
	year = {2020},
	note = {ISSN: 2226-0358},
	pages = {799--806},
}

@article{foglia_revealing_2019,
	title = {Revealing the competing contributions of charge carriers, excitons, and defects to the non-equilibrium optical properties of {ZnO}},
	volume = {6},
	issn = {2329-7778 (Print) 2329-7778 (Electronic) 2329-7778 (Linking)},
	doi = {10.1063/1.5088767},
	number = {3},
	journal = {Struct Dyn},
	author = {Foglia, L. and Vempati, S. and Tanda Bonkano, B. and Gierster, L. and Wolf, M. and Sadofev, S. and Stahler, J.},
	month = may,
	year = {2019},
	pmcid = {PMC6506340},
	pages = {034501},
}

@article{gerasimova_soft_2022,
	title = {The soft {X}-ray monochromator at the {SASE3} beamline of the {European} {XFEL}: from design to operation},
	volume = {29},
	issn = {1600-5775 (Electronic) 0909-0495 (Print) 0909-0495 (Linking)},
	doi = {10.1107/S1600577522007627},
	number = {Pt 5},
	journal = {J Synchrotron Radiat},
	author = {Gerasimova, N. and La Civita, D. and Samoylova, L. and Vannoni, M. and Villanueva, R. and Hickin, D. and Carley, R. and Gort, R. and Van Kuiken, B. E. and Miedema, P. and Le Guyarder, L. and Mercadier, L. and Mercurio, G. and Schlappa, J. and Teichman, M. and Yaroslavtsev, A. and Sinn, H. and Scherz, A.},
	month = sep,
	year = {2022},
	pmcid = {PMC9455211},
	pages = {1299--1308},
}

@article{gillmeister_ultrafast_2020,
	title = {Ultrafast coupled charge and spin dynamics in strongly correlated {NiO}},
	volume = {11},
	copyright = {2020 The Author(s)},
	issn = {2041-1723},
	url = {https://www.nature.com/articles/s41467-020-17925-8},
	doi = {10.1038/s41467-020-17925-8},
	language = {en},
	number = {1},
	urldate = {2024-08-26},
	journal = {Nature Communications},
	author = {Gillmeister, Konrad and Golež, Denis and Chiang, Cheng-Tien and Bittner, Nikolaj and Pavlyukh, Yaroslav and Berakdar, Jamal and Werner, Philipp and Widdra, Wolf},
	month = aug,
	year = {2020},
	note = {Publisher: Nature Publishing Group},
	pages = {4095},
}

@article{hennecke_angular_2019,
	title = {Angular {Momentum} {Flow} {During} {Ultrafast} {Demagnetization} of a {Ferrimagnet}},
	volume = {122},
	url = {https://link.aps.org/doi/10.1103/PhysRevLett.122.157202},
	doi = {10.1103/PhysRevLett.122.157202},
	number = {15},
	urldate = {2024-10-31},
	journal = {Physical Review Letters},
	author = {Hennecke, Martin and Radu, Ilie and Abrudan, Radu and Kachel, Torsten and Holldack, Karsten and Mitzner, Rolf and Tsukamoto, Arata and Eisebitt, Stefan},
	month = apr,
	year = {2019},
	note = {Publisher: American Physical Society},
	pages = {157202},
}

@article{holldack_femtospex_2014,
	title = {{FemtoSpeX}: a versatile optical pump-soft {X}-ray probe facility with 100 fs {X}-ray pulses of variable polarization},
	volume = {21},
	issn = {1600-5775 (Electronic) 0909-0495 (Linking)},
	doi = {10.1107/S1600577514012247},
	number = {Pt 5},
	journal = {J Synchrotron Radiat},
	author = {Holldack, K. and Bahrdt, J. and Balzer, A. and Bovensiepen, U. and Brzhezinskaya, M. and Erko, A. and Eschenlohr, A. and Follath, R. and Firsov, A. and Frentrup, W. and Le Guyader, L. and Kachel, T. and Kuske, P. and Mitzner, R. and Muller, R. and Pontius, N. and Quast, T. and Radu, I. and Schmidt, J. S. and Schussler-Langeheine, C. and Sperling, M. and Stamm, C. and Trabant, C. and Fohlisch, A.},
	month = sep,
	year = {2014},
	pages = {1090--104},
}

@article{ishikawa_compact_2012,
	title = {A compact {X}-ray free-electron laser emitting in the sub-ångström region},
	volume = {6},
	issn = {1749-4885 1749-4893},
	doi = {10.1038/nphoton.2012.141},
	number = {8},
	journal = {Nature Photonics},
	author = {Ishikawa, Tetsuya and Aoyagi, Hideki and Asaka, Takao and Asano, Yoshihiro and Azumi, Noriyoshi and Bizen, Teruhiko and Ego, Hiroyasu and Fukami, Kenji and Fukui, Toru and Furukawa, Yukito and Goto, Shunji and Hanaki, Hirofumi and Hara, Toru and Hasegawa, Teruaki and Hatsui, Takaki and Higashiya, Atsushi and Hirono, Toko and Hosoda, Naoyasu and Ishii, Miho and Inagaki, Takahiro and Inubushi, Yuichi and Itoga, Toshiro and Joti, Yasumasa and Kago, Masahiro and Kameshima, Takashi and Kimura, Hiroaki and Kirihara, Yoichi and Kiyomichi, Akio and Kobayashi, Toshiaki and Kondo, Chikara and Kudo, Togo and Maesaka, Hirokazu and Maréchal, Xavier M. and Masuda, Takemasa and Matsubara, Shinichi and Matsumoto, Takahiro and Matsushita, Tomohiro and Matsui, Sakuo and Nagasono, Mitsuru and Nariyama, Nobuteru and Ohashi, Haruhiko and Ohata, Toru and Ohshima, Takashi and Ono, Shun and Otake, Yuji and Saji, Choji and Sakurai, Tatsuyuki and Sato, Takahiro and Sawada, Kei and Seike, Takamitsu and Shirasawa, Katsutoshi and Sugimoto, Takashi and Suzuki, Shinsuke and Takahashi, Sunao and Takebe, Hideki and Takeshita, Kunikazu and Tamasaku, Kenji and Tanaka, Hitoshi and Tanaka, Ryotaro and Tanaka, Takashi and Togashi, Tadashi and Togawa, Kazuaki and Tokuhisa, Atsushi and Tomizawa, Hiromitsu and Tono, Kensuke and Wu, Shukui and Yabashi, Makina and Yamaga, Mitsuhiro and Yamashita, Akihiro and Yanagida, Kenichi and Zhang, Chao and Shintake, Tsumoru and Kitamura, Hideo and Kumagai, Noritaka},
	year = {2012},
	pages = {540--544},
}

@article{kachel_transient_2009,
	title = {Transient electronic and magnetic structures of nickel heated by ultrafast laser pulses},
	volume = {80},
	url = {https://link.aps.org/doi/10.1103/PhysRevB.80.092404},
	doi = {10.1103/PhysRevB.80.092404},
	number = {9},
	urldate = {2024-11-27},
	journal = {Physical Review B},
	author = {Kachel, T. and Pontius, N. and Stamm, C. and Wietstruk, M. and Aziz, E. F. and Dürr, H. A. and Eberhardt, W. and de Groot, F. M. F.},
	month = sep,
	year = {2009},
	note = {Publisher: American Physical Society},
	pages = {092404},
}

@article{ko_construction_2017,
	title = {Construction and {Commissioning} of {PAL}-{XFEL} {Facility}},
	volume = {7},
	copyright = {http://creativecommons.org/licenses/by/3.0/},
	issn = {2076-3417},
	url = {https://www.mdpi.com/2076-3417/7/5/479},
	doi = {10.3390/app7050479},
	language = {en},
	number = {5},
	urldate = {2024-11-27},
	journal = {Applied Sciences},
	author = {Ko, In Soo and Kang, Heung-Sik and Heo, Hoon and Kim, Changbum and Kim, Gyujin and Min, Chang-Ki and Yang, Haeryong and Baek, Soung Youl and Choi, Hyo-Jin and Mun, Geonyeong and Park, Byoung Ryul and Suh, Young Jin and Shin, Dong Cheol and Hu, Jinyul and Hong, Juho and Jung, Seonghoon and Kim, Sang-Hee and Kim, KwangHoon and Na, Donghyun and Park, Soung Soo and Park, Yong Jung and Jung, Young Gyu and Jeong, Seong Hun and Lee, Hong Gi and Lee, Sangbong and Lee, Sojeong and Oh, Bonggi and Suh, Hyung Suck and Han, Jang-Hui and Kim, Min Ho and Jung, Nam-Suk and Kim, Young-Chan and Lee, Mong-Soo and Lee, Bong-Ho and Sung, Chi-Won and Mok, Ik-Su and Yang, Jung-Moo and Parc, Yong Woon and Lee, Woul-Woo and Lee, Chae-Soon and Shin, Hocheol and Kim, Ji Hwa and Kim, Yongsam and Lee, Jae Hyuk and Park, Sang-Youn and Kim, Jangwoo and Park, Jaeku and Eom, Intae and Rah, Seungyu and Kim, Sunam and Nam, Ki Hyun and Park, Jaehyun and Park, Jaehun and Kim, Sangsoo and Kwon, Soonnam and An, Ran and Park, Sang Han and Kim, Kyung Sook and Hyun, Hyojung and Kim, Seung Nam and Kim, Seonghan and Yu, Chung-Jong and Kim, Bong-Soo and Kang, Tai-Hee and Kim, Kwang-Woo and Kim, Seung-Hwan and Lee, Hee-Seock and Lee, Heung-Soo and Park, Ki-Hyeon and Koo, Tae-Yeong and Kim, Dong-Eon and Lee, Ki Bong},
	month = may,
	year = {2017},
	note = {Number: 5
Publisher: Multidisciplinary Digital Publishing Institute},
	pages = {479},
}

@article{le_guyader_photon-shot-noise-limited_2023,
	title = {Photon-shot-noise-limited transient absorption soft {X}-ray spectroscopy at the {European} {XFEL}},
	volume = {30},
	issn = {1600-5775 (Electronic) 0909-0495 (Print) 0909-0495 (Linking)},
	doi = {10.1107/S1600577523000619},
	number = {Pt 2},
	journal = {J Synchrotron Radiat},
	author = {Le Guyader, L. and Eschenlohr, A. and Beye, M. and Schlotter, W. and Doring, F. and Carinan, C. and Hickin, D. and Agarwal, N. and Boeglin, C. and Bovensiepen, U. and Buck, J. and Carley, R. and Castoldi, A. and D'Elia, A. and Delitz, J. T. and Ehsan, W. and Engel, R. and Erdinger, F. and Fangohr, H. and Fischer, P. and Fiorini, C. and Fohlisch, A. and Gelisio, L. and Gensch, M. and Gerasimova, N. and Gort, R. and Hansen, K. and Hauf, S. and Izquierdo, M. and Jal, E. and Kamil, E. and Karabekyan, S. and Kluyver, T. and Laarmann, T. and Lojewski, T. and Lomidze, D. and Maffessanti, S. and Mamyrbayev, T. and Marcelli, A. and Mercadier, L. and Mercurio, G. and Miedema, P. S. and Ollefs, K. and Rossnagel, K. and Rosner, B. and Rothenbach, N. and Samartsev, A. and Schlappa, J. and Setoodehnia, K. and Sorin Chiuzbaian, G. and Spieker, L. and Stamm, C. and Stellato, F. and Techert, S. and Teichmann, M. and Turcato, M. and Van Kuiken, B. and Wende, H. and Yaroslavtsev, A. and Zhu, J. and Molodtsov, S. and David, C. and Porro, M. and Scherz, A.},
	month = mar,
	year = {2023},
	pmcid = {PMC10000791},
	pages = {284--300},
}

@misc{noauthor_Ni_Data_2020,
	title = {Data recorded in the experiments on Ni-metal at the European XFEL are available at DOI: 10.22003/XFEL.EU-DATA-002161-00},
	doi = {10.22003/XFEL.EU-DATA-002161-00},
    author = {Lojewski, Tobias and Elhanoty, Mohamed F. and Le Guyader, Loïc and Grånäs, Oscar and Agarwal, Naman and Boeglin, Christine and Carley, Robert and Castoldi, Andrea and David, Christian and Deiter, Carsten and Döring, Florian and Engel, RobinY and Erdinger, Florian and Fangohr, Hans and Fiorini, Carlo and Fischer, Peter and Gerasimova, Natalia and Gort, Rafael and deGroot, Frank and Hansen, Karsten and Hauf, Steffen and Hickin, David and Izquierdo, Manuel and Van Kuiken, Benjamin E. and Kvashnin, Yaroslav and Lambert, Charles-Henri and Lomidze, David and Maffessanti, Stefano and Mercadier, Laurent and Mercurio, Giuseppe and Miedema, Piter S. and Ollefs, Katharina and Pace, Matthias and Porro, Matteo and Rezvani, Javad and Rösner, Benedikt and Rothenbach, Nico and Samartsev, Andrey and Scherz, Andreas and Schlappa, Justina and Stamm, Christian and Teichmann, Martin and Thunström, Patrik and Turcato, Monica and Yaroslavtsev, Alexander and Zhu, Jun and Beye, Martin and Wende, Heiko and Bovensiepen, Uwe and Eriksson, Olle and Eschenlohr, Andrea},
    year = {2020},
}

@misc{noauthor_NiO_Data_2021,
	title = {Data recorded in the experiments on NiO at the European XFEL are available at DOI: 10.22003/xfel.eu-data-002589-00},
	doi = {10.22003/xfel.eu-data-002589-00},
    author = {Lojewski, Tobias and Golez, Denis and Ollefs, Katharina and Guyader, Loïc Le and Kämmerer, Lea and Rothenbach, Nico and Engel, Robin Y. and Miedema, Piter S. and Beye, Martin and Chiuzbăian, Gheorghe S. and Carley, Robert and Gort, Rafael and Van Kuiken, Benjamin E. and Mercurio, Giuseppe and Schlappa, Justina and Yaroslavtsev, Alexander and Scherz, Andreas and Döring, Florian and David, Christian and Wende, Heiko and Bovensiepen, Uwe and Eckstein, Martin and Werner, Philipp and Eschenlohr, Andrea},
    year = {2021},
}

@article{lojewski_interplay_2023,
	title = {The interplay of local electron correlations and ultrafast spin dynamics in fcc {Ni}},
	volume = {11},
	issn = {2166-3831},
	doi = {10.1080/21663831.2023.2210606},
	number = {8},
	journal = {Materials Research Letters},
	author = {Lojewski, Tobias and Elhanoty, Mohamed F. and Le Guyader, Loïc and Grånäs, Oscar and Agarwal, Naman and Boeglin, Christine and Carley, Robert and Castoldi, Andrea and David, Christian and Deiter, Carsten and Döring, Florian and Engel, RobinY and Erdinger, Florian and Fangohr, Hans and Fiorini, Carlo and Fischer, Peter and Gerasimova, Natalia and Gort, Rafael and deGroot, Frank and Hansen, Karsten and Hauf, Steffen and Hickin, David and Izquierdo, Manuel and Van Kuiken, Benjamin E. and Kvashnin, Yaroslav and Lambert, Charles-Henri and Lomidze, David and Maffessanti, Stefano and Mercadier, Laurent and Mercurio, Giuseppe and Miedema, Piter S. and Ollefs, Katharina and Pace, Matthias and Porro, Matteo and Rezvani, Javad and Rösner, Benedikt and Rothenbach, Nico and Samartsev, Andrey and Scherz, Andreas and Schlappa, Justina and Stamm, Christian and Teichmann, Martin and Thunström, Patrik and Turcato, Monica and Yaroslavtsev, Alexander and Zhu, Jun and Beye, Martin and Wende, Heiko and Bovensiepen, Uwe and Eriksson, Olle and Eschenlohr, Andrea},
	year = {2023},
	pages = {655--661},
}

@misc{lojewski_photo-induced_2024,
	title = {Photo-induced charge-transfer renormalization in {NiO}},
	url = {http://arxiv.org/abs/2305.10145},
	doi = {10.48550/arXiv.2305.10145},
	urldate = {2024-08-26},
	publisher = {arXiv},
	author = {Lojewski, Tobias and Golez, Denis and Ollefs, Katharina and Guyader, Loïc Le and Kämmerer, Lea and Rothenbach, Nico and Engel, Robin Y. and Miedema, Piter S. and Beye, Martin and Chiuzbăian, Gheorghe S. and Carley, Robert and Gort, Rafael and Van Kuiken, Benjamin E. and Mercurio, Giuseppe and Schlappa, Justina and Yaroslavtsev, Alexander and Scherz, Andreas and Döring, Florian and David, Christian and Wende, Heiko and Bovensiepen, Uwe and Eckstein, Martin and Werner, Philipp and Eschenlohr, Andrea},
	month = may,
	year = {2024},
	note = {arXiv:2305.10145 [cond-mat]},
}

@article{lopez-flores_time-resolved_2012,
	title = {Time-resolved x-ray magnetic circular dichroism study of ultrafast demagnetization in a {CoPd} ferromagnetic film excited by circularly polarized laser pulse},
	volume = {86},
	url = {https://link.aps.org/doi/10.1103/PhysRevB.86.014424},
	doi = {10.1103/PhysRevB.86.014424},
	number = {1},
	urldate = {2024-10-31},
	journal = {Physical Review B},
	author = {López-Flores, Víctor and Arabski, Jacek and Stamm, Christian and Halté, Valérie and Pontius, Niko and Beaurepaire, Eric and Boeglin, Christine},
	month = jul,
	year = {2012},
	note = {Publisher: American Physical Society},
	pages = {014424},
}

@article{lopez-flores_role_2013,
	title = {Role of critical spin fluctuations in ultrafast demagnetization of transition-metal rare-earth alloys},
	volume = {87},
	url = {https://link.aps.org/doi/10.1103/PhysRevB.87.214412},
	doi = {10.1103/PhysRevB.87.214412},
	number = {21},
	urldate = {2024-10-31},
	journal = {Physical Review B},
	author = {López-Flores, V. and Bergeard, N. and Halté, V. and Stamm, C. and Pontius, N. and Hehn, M. and Otero, E. and Beaurepaire, E. and Boeglin, C.},
	month = jun,
	year = {2013},
	note = {Publisher: American Physical Society},
	pages = {214412},
}

@article{patterson_coherent_2010,
	title = {Coherent science at the {SwissFEL} x-ray laser},
	volume = {12},
	url = {https://dx.doi.org/10.1088/1367-2630/12/3/035012},
	doi = {10.1088/1367-2630/12/3/035012},
	number = {3},
	journal = {New Journal of Physics},
	author = {Patterson, B. D. and Abela, R. and Braun, H.-H. and Flechsig, U. and Ganter, R. and Kim, Y. and Kirk, E. and Oppelt, A. and Pedrozzi, M. and Reiche, S. and Rivkin, L. and Schmidt, Th and Schmitt, B. and Strocov, V. N. and Tsujino, S. and Wrulich, A. F.},
	month = mar,
	year = {2010},
	pages = {035012},
}

@article{pergament_versatile_2016,
	title = {Versatile optical laser system for experiments at the {European} {X}-ray free-electron laser facility},
	volume = {24},
	issn = {1094-4087},
	doi = {10.1364/OE.24.029349},
	language = {eng},
	number = {26},
	journal = {Optics Express},
	author = {Pergament, M. and Palmer, G. and Kellert, M. and Kruse, K. and Wang, J. and Wissmann, L. and Wegner, U. and Emons, M. and Kane, D. and Priebe, G. and Venkatesan, S. and Jezynski, T. and Pallas, F. and Lederer, M. J.},
	month = dec,
	year = {2016},
	pmid = {28059324},
	pages = {29349--29359},
}

@book{philip_willmott_introduction_2019,
	title = {An {Introduction} to {Synchrotron} {Radiation}: {Techniques} and {Applications}},
	isbn = {978-1-119-28045-3},
	url = {https://doi.org/10.1002/9781119280453},
	language = {en},
	urldate = {2024-09-01},
	publisher = {John Wiley \& Sons, Ltd},
	author = {Willmott, Philip},
	month = mar,
	year = {2019},
	doi = {10.1002/9781119280453},
	note = {\_eprint: https://onlinelibrary.wiley.com/doi/epub/10.1002/9781119280453},
}

@article{porro_minisdd-based_2021,
	title = {The {MiniSDD}-{Based} 1-{Mpixel} {Camera} of the {DSSC} {Project} for the {European} {XFEL}},
	volume = {68},
	issn = {1558-1578},
	url = {https://ieeexplore.ieee.org/document/9419081},
	doi = {10.1109/TNS.2021.3076602},
	number = {6},
	urldate = {2024-11-11},
	journal = {IEEE Transactions on Nuclear Science},
	author = {Porro, Matteo and Andricek, Ladislav and Aschauer, Stefan and Castoldi, Andrea and Donato, Mattia and Engelke, Jan and Erdinger, Florian and Fiorini, Carlo and Fischer, Peter and Graafsma, Heinz and Grande, Andrea and Guazzoni, Chiara and Hansen, Karsten and Hauf, Steffen and Kalavakuru, Pradeep and Klaer, Helmut and Tangl, Manfred and Kugel, Andreas and Kuster, Markus and Lechner, Peter and Lomidze, David and Maffessanti, Stefano and Manghisoni, Massimo and Nidhi, Sneha and Okrent, Frank and Re, Valerio and Reckleben, Christian and Riceputi, Elisa and Richter, Rainer and Samartsev, Andrey and Schlee, Stephan and Soldat, Jan and Strüder, Lothar and Szymanski, Janusz and Turcato, Monica and Weidenspointner, Georg and Wunderer, Cornelia B.},
	month = jun,
	year = {2021},
	note = {Conference Name: IEEE Transactions on Nuclear Science},
	pages = {1334--1350},
}

@inproceedings{raubenheimer_lcls-ii-he_2018,
	address = {Geneva, Switzerland},
	series = {{ICFA} {Advanced} {Beam} {Dynamics} {Workshop}},
	title = {The {LCLS}-{II}-{HE}, {A} {High} {Energy} {Upgrade} of the {LCLS}-{II}},
	isbn = {978-3-95450-206-6},
	url = {http://jacow.org/fls2018/papers/mop1wa02.pdf},
	doi = {doi:10.18429/JACoW-FLS2018-MOP1WA02},
	language = {english},
	booktitle = {Proc. 60th {ICFA} {Advanced} {Beam} {Dynamics} {Workshop} ({FLS}'18), {Shanghai}, {China}, 5-9 {March} 2018},
	publisher = {JACoW Publishing},
	author = {Raubenheimer, T. O.},
	month = jun,
	year = {2018},
	note = {Issue: 60},
	pages = {6--11},
}

@article{rothenbach_microscopic_2019,
	title = {Microscopic nonequilibrium energy transfer dynamics in a photoexcited metal/insulator heterostructure},
	volume = {100},
	url = {https://link.aps.org/doi/10.1103/PhysRevB.100.174301},
	doi = {10.1103/PhysRevB.100.174301},
	number = {17},
	urldate = {2024-11-27},
	journal = {Physical Review B},
	author = {Rothenbach, N. and Gruner, M. E. and Ollefs, K. and Schmitz-Antoniak, C. and Salamon, S. and Zhou, P. and Li, R. and Mo, M. and Park, S. and Shen, X. and Weathersby, S. and Yang, J. and Wang, X. J. and Pentcheva, R. and Wende, H. and Bovensiepen, U. and Sokolowski-Tinten, K. and Eschenlohr, A.},
	month = nov,
	year = {2019},
	note = {Publisher: American Physical Society},
	pages = {174301},
}

@article{rothenbach_effect_2021,
	title = {Effect of lattice excitations on transient near-edge x-ray absorption spectroscopy},
	volume = {104},
	url = {https://link.aps.org/doi/10.1103/PhysRevB.104.144302},
	doi = {10.1103/PhysRevB.104.144302},
	number = {14},
	urldate = {2024-11-27},
	journal = {Physical Review B},
	author = {Rothenbach, N. and Gruner, M. E. and Ollefs, K. and Schmitz-Antoniak, C. and Salamon, S. and Zhou, P. and Li, R. and Mo, M. and Park, S. and Shen, X. and Weathersby, S. and Yang, J. and Wang, X. J. and Šipr, O. and Ebert, H. and Sokolowski-Tinten, K. and Pentcheva, R. and Bovensiepen, U. and Eschenlohr, A. and Wende, H.},
	month = oct,
	year = {2021},
	note = {Publisher: American Physical Society},
	pages = {144302},
}

@article{santomauro_localized_2017,
	title = {Localized holes and delocalized electrons in photoexcited inorganic perovskites: {Watching} each atomic actor by picosecond {X}-ray absorption spectroscopy},
	volume = {4},
	issn = {2329-7778},
	shorttitle = {Localized holes and delocalized electrons in photoexcited inorganic perovskites},
	url = {https://doi.org/10.1063/1.4971999},
	doi = {10.1063/1.4971999},
	number = {4},
	urldate = {2024-11-27},
	journal = {Structural Dynamics},
	author = {Santomauro, Fabio G. and Grilj, Jakob and Mewes, Lars and Nedelcu, Georgian and Yakunin, Sergii and Rossi, Thomas and Capano, Gloria and Al Haddad, André and Budarz, James and Kinschel, Dominik and Ferreira, Dario S. and Rossi, Giacomo and Gutierrez Tovar, Mario and Grolimund, Daniel and Samson, Valerie and Nachtegaal, Maarten and Smolentsev, Grigory and Kovalenko, Maksym V. and Chergui, Majed},
	month = dec,
	year = {2016},
	pages = {044002},
}

@misc{noauthor_scs_2024,
	title = {{SCS} / {ToolBox} · {GitLab}},
	url = {https://git.xfel.eu/SCS/ToolBox},
	language = {en},
	urldate = {2024-08-26},
	journal = {GitLab},
	month = aug,
	year = {2024},
    author = {SCS-team},
}

@article{stamm_femtosecond_2007,
	title = {Femtosecond modification of electron localization and transfer of angular momentum in nickel},
	volume = {6},
	issn = {1476-1122 (Print) 1476-1122 (Linking)},
	doi = {10.1038/nmat1985},
	number = {10},
	journal = {Nat Mater},
	author = {Stamm, C. and Kachel, T. and Pontius, N. and Mitzner, R. and Quast, T. and Holldack, K. and Khan, S. and Lupulescu, C. and Aziz, E. F. and Wietstruk, M. and Durr, H. A. and Eberhardt, W.},
	month = oct,
	year = {2007},
	pages = {740--3},
}

@article{stamm_femtosecond_2010,
	title = {Femtosecond x-ray absorption spectroscopy of spin and orbital angular momentum in photoexcited {Ni} films during ultrafast demagnetization},
	volume = {81},
	url = {https://link.aps.org/doi/10.1103/PhysRevB.81.104425},
	doi = {10.1103/PhysRevB.81.104425},
	number = {10},
	urldate = {2024-11-27},
	journal = {Physical Review B},
	author = {Stamm, C. and Pontius, N. and Kachel, T. and Wietstruk, M. and Dürr, H. A.},
	month = mar,
	year = {2010},
	note = {Publisher: American Physical Society},
	pages = {104425},
}

@article{stamm_x-ray_2020,
	title = {X-ray detection of ultrashort spin current pulses in synthetic antiferromagnets},
	volume = {127},
	issn = {0021-8979},
	url = {https://doi.org/10.1063/5.0006095},
	doi = {10.1063/5.0006095},
	number = {22},
	urldate = {2024-11-01},
	journal = {Journal of Applied Physics},
	author = {Stamm, C. and Murer, C. and Wörnle, M. S. and Acremann, Y. and Gort, R. and Däster, S. and Reid, A. H. and Higley, D. J. and Wandel, S. F. and Schlotter, W. F. and Gambardella, P.},
	month = jun,
	year = {2020},
	pages = {223902},
}

@article{stepanov_short_2016,
	title = {Short {X}-ray pulses from third-generation light sources},
	volume = {23},
	issn = {1600-5775 (Electronic) 0909-0495 (Linking)},
	doi = {10.1107/S1600577515019281},
	number = {1},
	journal = {J Synchrotron Radiat},
	author = {Stepanov, A. G. and Hauri, C. P.},
	month = jan,
	year = {2016},
	pages = {141--51},
}

@book{stohr_magnetism_2006,
	address = {Berlin, Heidelberg},
	title = {Magnetism},
	copyright = {http://www.springer.com/tdm},
	isbn = {978-3-540-30282-7},
	url = {http://link.springer.com/10.1007/978-3-540-30283-4},
	language = {en},
	urldate = {2024-09-01},
	publisher = {Springer},
	author = {Stöhr, Joachim and Siegmann, Hans Christoph},
	year = {2006},
	doi = {10.1007/978-3-540-30283-4},
}

@article{tschentscher_photon_2017,
	title = {Photon {Beam} {Transport} and {Scientific} {Instruments} at the {European} {XFEL}},
	volume = {7},
	copyright = {http://creativecommons.org/licenses/by/3.0/},
	issn = {2076-3417},
	url = {https://www.mdpi.com/2076-3417/7/6/592},
	doi = {10.3390/app7060592},
	language = {en},
	number = {6},
	urldate = {2024-11-27},
	journal = {Applied Sciences},
	author = {Tschentscher, Thomas and Bressler, Christian and Grünert, Jan and Madsen, Anders and Mancuso, Adrian P. and Meyer, Michael and Scherz, Andreas and Sinn, Harald and Zastrau, Ulf},
	month = jun,
	year = {2017},
	note = {Number: 6
Publisher: Multidisciplinary Digital Publishing Institute},
	pages = {592},
}

@article{uemura_femtosecond_2021,
	title = {Femtosecond {Charge} {Density} {Modulations} in {Photoexcited} {CuWO4}},
	volume = {125},
	issn = {1932-7447},
	url = {https://doi.org/10.1021/acs.jpcc.0c10525},
	doi = {10.1021/acs.jpcc.0c10525},
	number = {13},
	urldate = {2024-11-25},
	journal = {The Journal of Physical Chemistry C},
	author = {Uemura, Yohei and Ismail, Ahmed S. M. and Park, Sang Han and Kwon, Soonnam and Kim, Minseok and Niwa, Yasuhiro and Wadati, Hiroki and Elnaggar, Hebatalla and Frati, Federica and Haarman, Ties and Höppel, Niko and Huse, Nils and Hirata, Yasuyuki and Zhang, Yujun and Yamagami, Kohei and Yamamoto, Susumu and Matsuda, Iwao and Katayama, Tetsuo and Togashi, Tadashi and Owada, Shigeki and Yabashi, Makina and Halisdemir, Uufuk and Koster, Gertjan and Yokoyama, Toshihiko and Weckhuysen, Bert M. and de Groot, Frank M. F.},
	month = apr,
	year = {2021},
	note = {Publisher: American Chemical Society},
	pages = {7329--7336},
}

@article{uemura_hole_2022,
	title = {Hole {Dynamics} in {Photoexcited} {Hematite} {Studied} with {Femtosecond} {Oxygen} {K}‑edge {X}‑ray {Absorption} {Spectroscopy}},
	volume = {13},
	url = {https://tohoku.elsevierpure.com/en/publications/hole-dynamics-in-photoexcited-hematite-studied-with-femtosecond-o},
	doi = {10.1021/acs.jpclett.2c00295},
	language = {English},
	number = {19},
	urldate = {2024-11-26},
	journal = {Journal of Physical Chemistry Letters},
	author = {Uemura, Yohei and Ismail, Ahmed S. M. and Park, Sang Han and Kwon, Soonnam and Kim, Minseok and Elnaggar, Hebatalla and Frati, Federica and Wadati, Hiroki and Hirata, Yasuyuki and Zhang, Yujun and Yamagami, Kohei and Yamamoto, Susumu and Matsuda, Iwao and Halisdemir, Ufuk and Koster, Gertjan and Milne, Christopher and Ammann, Markus and Weckhuysen, Bert M. and Groot, Frank M. F. de},
	year = {2022},
	pmid = {35512383},
	note = {Publisher: American Chemical Society},
	pages = {4207--4214},
}

@article{wang_l-edge_2018,
	title = {L-edge sum rule analysis on 3d transition metal sites: from d 10 to d 0 and towards application to extremely dilute metallo-enzymes},
	volume = {20},
	shorttitle = {L-edge sum rule analysis on 3d transition metal sites},
	url = {https://pubs.rsc.org/en/content/articlelanding/2018/cp/c7cp06624d},
	doi = {10.1039/C7CP06624D},
	language = {en},
	number = {12},
	urldate = {2024-08-26},
	journal = {Physical Chemistry Chemical Physics},
	author = {Wang, Hongxin and Friedrich, Stephan and Li, Lei and Mao, Ziliang and Ge, Pinghua and Balasubramanian, Mahalingam and S. Patil, Daulat},
	year = {2018},
	note = {Publisher: Royal Society of Chemistry},
	pages = {8166--8176},
}

@article{yu_charge_2012,
	title = {Charge and {Ion} {Transport} in {NiO} and {Aspects} of {Ni} {Oxidation} from {First} {Principles}},
	volume = {116},
	issn = {1932-7447},
	url = {https://doi.org/10.1021/jp208080v},
	doi = {10.1021/jp208080v},
	number = {2},
	urldate = {2024-11-27},
	journal = {The Journal of Physical Chemistry C},
	author = {Yu, Jianguo and Rosso, Kevin M. and Bruemmer, Stephen M.},
	month = jan,
	year = {2012},
	note = {Publisher: American Chemical Society},
	pages = {1948--1954},
}

@article{zhang_stable_2023,
	title = {Stable {Mott} {Polaron} {State} {Limits} the {Charge} {Density} in {Lead} {Halide} {Perovskites}},
	volume = {8},
	issn = {2380-8195 (Print) 2380-8195 (Electronic)},
	doi = {10.1021/acsenergylett.2c01949},
	number = {1},
	journal = {ACS Energy Lett},
	author = {Zhang, H. and Debroye, E. and Vina-Bausa, B. and Valli, D. and Fu, S. and Zheng, W. and Di Virgilio, L. and Gao, L. and Frost, J. M. and Walsh, A. and Hofkens, J. and Wang, H. I. and Bonn, M.},
	month = jan,
	year = {2023},
	pmcid = {PMC9841606},
	pages = {420--428},
}
\end{document}